\documentclass[10pt,sigplan,screen]{acmart}
\AtBeginDocument{%
  \providecommand\BibTeX{{%
      \normalfont B\kern-0.5em{\scshape i\kern-0.25em b}\kern-0.8em\TeX}}}

\usepackage{ifluatex,ifxetex}
\ifxetex
\usepackage{fontspec}
\usepackage[utf8]{inputenc}
\fi
\ifluatex
\usepackage{fontspec}
\usepackage[utf8]{inputenc}
\fi

\usepackage{microtype} 
\usepackage{balance} 

\usepackage{algpseudocode}
\usepackage[linesnumbered,ruled,vlined,commentsnumbered]{algorithm2e}
\SetAlFnt{\footnotesize}
\algnewcommand{\algorithmicforeach}{\textbf{for each}}
\algdef{SE}[FOR]{ForEach}{EndForEach}[1]
{\algorithmicforeach\ #1\ \algorithmicdo}
{\algorithmicend\ \algorithmicforeach}

\usepackage{import}
\usepackage[subpreambles=true]{standalone}
\usepackage{siunitx}
\usepackage{amsmath,amsfonts}
\usepackage{float,here}

\usepackage[frozencache]{minted}
\setminted[python]{
  fontsize=\scriptsize,
  autogobble,
  escapeinside=||,
  style=bw,
}
\setminted[ocaml]{
  fontsize=\scriptsize,
  autogobble,
  escapeinside=||,
  style=bw
}
\setminted[text]{
  fontsize=\scriptsize,
  autogobble,
  escapeinside=||
}

\usepackage{listings}
\usepackage{color}
\usepackage{xcolor}
\definecolor{Green}{rgb}{0,0.6,0}
\definecolor{Gray}{rgb}{0.5,0.5,0.5}
\definecolor{Purple}{rgb}{0.58,0,0.82}
\definecolor{backgroundColour}{rgb}{0.95,0.95,0.92}

\lstdefinestyle{defhighlight} {
  commentstyle=\color{Green},
  keywordstyle=\color{magenta},
  numberstyle=\tiny\color{Gray},
  stringstyle=\color{Purple},
  basicstyle=\scriptsize\ttfamily,
  breakatwhitespace=false,
  breaklines=true,
  captionpos=b,
  keepspaces=true,
  numbers=left,
  numbersep=5pt,
  showspaces=false,
  showstringspaces=false,
  showtabs=false,
  tabsize=2,
}
\lstdefinestyle{c}{ style = defhighlight, language = C }
\lstdefinestyle{python}{ style = defhighlight, language = Python }
\lstdefinestyle{caml}{ style = defhighlight, language = Caml }

\usepackage{multicol}
\usepackage{lipsum}

\usepackage{graphicx}\graphicspath{ {./figs/} }

\usepackage{tikz}
\usepackage{smartdiagram}
\usepackage{pgfplots}
\usepackage{subcaption}

\usetikzlibrary{decorations.pathmorphing}
\usetikzlibrary{patterns}

\usepackage{todonotes}

\newcommand{\jit}{JIT}
\newcommand{\aot}{AOT}
\newcommand{\astree}{AST}
\newcommand{\baccaml}{BacCaml}
\newcommand{\mincaml}{MinCaml}
\newcommand{\rpython}{RPython}
\newcommand{\truffle}{Truffle}
\newcommand{\vm}{VM}

\newcommand{\sh}{Stack Hybridization}

\newcommand{\minus}{\scalebox{0.75}[1.0]{$-$}}
\newcommand{\mincamlmm}{MinCaml$\minus$$\minus$}

\setcopyright{acmcopyright}
\acmPrice{15.00}
\acmDOI{10.1145/3426422.3426977}
\acmYear{2020}
\copyrightyear{2020}
\acmSubmissionID{dls20main-p4-p}
\acmISBN{978-1-4503-8175-8/20/11}
\acmConference[DLS '20]{Proceedings of the 16th ACM SIGPLAN International Symposium on Dynamic Languages}{November 17, 2020}{Virtual, USA}
\acmBooktitle{Proceedings of the 16th ACM SIGPLAN International Symposium on Dynamic Languages (DLS '20), November 17, 2020, Virtual, USA}

\begin{document}

\title{Amalgamating Different JIT Compilations in a Meta-tracing JIT Compiler Framework}

\author{Yusuke Izawa}
\email{izawa@prg.is.titech.ac.jp}
\affiliation{%
  \institution{Tokyo Institute of Technology}
  \state{Tokyo}
  \country{Japan}
}
\author{Hidehiko Masuhara}
\email{masuhara@acm.org}
\affiliation{%
  \institution{Tokyo Institute of Technology}
  \state{Tokyo}
  \country{Japan}
}


\begin{abstract}
  Most virtual machines employ just-in-time (JIT) compilers to achieve
  high-performance. Trace-based compilation and method-based compilation are two
  major compilation strategies in JIT compilers. In general, the former excels
  in compiling programs with more in-depth method calls and more dynamic
  branches, while the latter is suitable for a wide range of programs. Some
  previous studies have suggested that each strategy has its advantages and
  disadvantages,and there is no clear winner.

  In this paper, we present a new approach, namely, the meta-hybrid JIT compilation
  strategy, mixing the two strategies in a meta-tracing JIT compiler. As a prototype,
  we implemented a simple meta-tracing JIT compiler framework called BacCaml based on
  the MinCaml compiler by following RPython’s architecture. We also report that some
  programs ran faster by the hybrid compilation in our experiments.
\end{abstract}

\ccsdesc[500]{Software and its engineering~Just-in-time compilers}

\keywords{JIT compiler, language implementation framework, meta-tracing JIT
  compiler, RPython}

\maketitle

\section{Introduction}
\label{sec:introduction}

Just-in-Time (\jit{}) compilation is widely used in modern programming language
implementations that include
Java~\cite{paleczny2001java,Gal:2006:10.1145/1134760.1134780},
JavaScript~\cite{Gal2009,ionmonkey,googlev8}, and PHP~\cite{Adams2014,
  10.1145/3192366.3192374} to name a few. There is a variety of JIT
compilers that commonly identify ``hot spots'' or frequently-executed parts of a
program at runtime, collect runtime information, and generate machine code.
By compiling only frequently-executed parts of a program, the compilation is
fast enough to be executed at runtime. By exploiting runtime information,
compilers can execute aggressive optimization techniques such as inlining,
type specialization and produce as efficient---sometimes more efficient---code
as the that code generated with traditional static compilers.

\jit{} compilers can be classified by the strategy of selecting compilation
targets, namely method-based and trace-based strategies.  The method-based strategy
involves using methods (or functions) as a unit of compilation, which has been used
in many \jit{} compilers~\cite{Smalltalk80:10.1145/800017.800542,
  SELF:10.1145/38807.38828,paleczny2001java}.
It shares many optimization techniques with traditional static compilers.
The trace-based strategy involves using a \emph{trace} of a program execution, which
is a sequence of instructions during a particular run of a program, as a compilation
unit~\cite{Bala2000,Bebenita2010,Bolz2009,Gal2009}. It effectively executes
inlining, loop unrolling and type specialization.

Neither of the two strategies is better than the other; instead, each
works better for different types of programs. While the method-based strategy
works well on average, the trace-based strategy exhibits polarized performance
characteristics. It works better for programs with many biased conditional
branches and deeply nested function
calls~\cite{Bala2000,Bolz2009,Gal2009,Inoue2011}. However, this strategy can cause
severe overheads when the path of an execution varies, which is known as the
\emph{path-divergence problem}~\cite{Hayashizaki:2011:IPT:1950365.1950412,
  Huang2016}.

It is seems straightforward to combine the two strategies, i.e., when compiling a
different part of a program, using a strategy that works better for that part.
However, the following questions need to be answered: (1) how can we construct such a
compilation engine without actually creating two very different compilers, (2) how do
the code fragments that are compiled by different strategies interact with each
other, and (3) how can we determine a compilation strategy to compile a part of a
program.

There are not many studies on these regards. To the best of our knowledge,
only region-based compilation by HipHop Virtual Machine
(HHVM)~\cite{10.1145/3192366.3192374} and lazy basic block versioning by
Higgs~\cite{ChevalierBoisvert_et_al:LIPIcs:2015:5219} are the strategies that
supports both strategies. HHVM is designed for PHP and Hack, and implemented by
making a method-based compiler more flexible in selecting compilation targets. Lazy
basic block versioning is provided by Higgs JavaScript \vm{}. It is a \jit{} code
generation technique working at the level of basic blocks, defining a basic block as
a single-entry and single-exit sequence of instructions, and performs both strategies
by combining type-specialized basic blocks.

We propose a \emph{meta-hybrid \jit{} compiler framework} to combine trace- and
method-based strategies, to study questions (1) and (2), while leaving question (3)
for future work. One major difference from the existing hybrid compilers is that we
design a \emph{meta-}\jit{} compiler framework, which can generate a \jit{} compiler
for a new language by merely writing an interpreter of the language.  Moreover, we
design it by extending a meta-\emph{tracing} compilation framework by following the
\rpython{}'s~\cite{Bolz2009, 10.1145/2991041.2991062} architecture.

We implemented \baccaml{}, a prototype implementation of our framework.
Though its architecture is based on the \rpython{}'s, it is a completely
different and much simpler implementation written in OCaml.
We modified the \mincaml{} compiler~\cite{Sumii2005} as the compiler backend.



\paragraph{Contributions and Organization}

This paper has the following contributions.

\begin{itemize}
\item We propose a meta-hybrid \jit{} compiler framework that
  can apply both trace- and method-based compilation strategies, i.e., hybrid
  compilation strategy, to different parts of a program under a meta-\jit{} compiler
  framework.
\item We present a technique to achieve method-based compilation with a
  meta-tracing \jit{} compiler by controlling the trace-based compiler to cover
  all the paths in a method.
\item We identify the problem of interconnecting code fragments compiled by the
  different strategies, and present a solution that involves dynamically switching
  the usage of call stacks.
\item We implemented a prototype of our framework called \baccaml{}, and confirmed
  that there are programs where the hybrid compilation strategy for our framework
  performs better.
\end{itemize}

The rest of this paper is organized as follows. In Section~\ref{sec:background},
we give an overview of method \jit{}, tracing \jit{}, and meta-tracing \jit{}
compilation which serve as the underlying techniques of our framework.
In Section~\ref{sec:architecture_overview}, we present our hybrid compiler
framework after discussing the advantages and disadvantages of tracing- and
method-based strategies.
In Section~\ref{sec:mixing_the_two_compilationss}, we introduce a technique
that enables method-based compilation with a meta-tracing \jit{}
compiler. Afterwards, we explain the problem and solution of combining code fragments
compiled by two compilation strategies.
In Section~\ref{sec:evaluation}, we evaluate the basic performance of the current
\baccaml{} implementation.
In Section~\ref{sec:experiment}, we report on a synthetic experiment in order to
confirm usefulness of the hybrid strategy.
In Section~\ref{sec:related_work}, we discusses related work.
Finally, we conclude the paper in Section~\ref{sec:conclusion_futurework}.

\section{Background}
\label{sec:background}

Before presenting the concept and implementation of our meta-hybrid \jit{} compiler
framework, we briefly review the following compilation techniques which are essential
of our framework; method-based compilation, trace-based compilation, and meta-tracing
\jit{} compilation.

\subsection{Method-Based JIT Compilation Technique}
\label{sec:method_jit_compilation}

\jit{} compilation is a commonly used technique in VM-based languages, including
Smalltalk-80~\cite{Smalltalk80:10.1145/800017.800542},
\textsc{Self}~\cite{SELF:10.1145/38807.38828} and
Java~\cite{IBM:10.1145/353171.353175, paleczny2001java}.
It performs the same compilation processes as the-back-end of the
ahead-of-time compilers but does only to a small set of methods during an execution of
a program. The compiled methods are ``hot spots'' in the execution as they are
usually chosen from frequently-executed ones by taking an execution profile. In
addition to the standard optimization techniques that are developed for the
ahead-of-time compilers, it performs more optimization techniques by exploiting
profiling information. One of those techniques is \emph{aggressive inlining}, which
selectively chooses inline method bodies only to the ones frequently executed. By not
inlining methods that are rarely executed, it can inline more nested method calls.

\subsection{Trace-Based JIT Compilation Technique}
\label{sec:tracing_jit_compilation}

Tracing optimization was initially investigated by the Dynamo
project~\cite{Bala2000} and was adopted for implementing compilers for many
languages such as Lua~\cite{luajit}, JavaScript~\cite{Gal2009}, Java
trace-\jit{}~\cite{Gal:2006:10.1145/1134760.1134780,Inoue2011} and the SPUR
project~\cite{Bebenita2010}.

Tracing \jit{} compilers track the execution of a program and generate a machine
code with hot paths. They convert a sequential code path called \emph{trace}
into native code while interpreting others~\cite{BOLZ2015408}. ``Trace'' is a
straight-line program; therefore, every possible branch is selected as the
actually-executed one. To ensure that the tracing and execution condition is the
same, a \textit{guard} code is placed at every possible point (e.g., if statements)
that go in another direction. The guard code determines whether the original
condition is still valid. If the condition is false, the machine code's execution
stops and falls back to the interpreter.


\subsection{Meta-Tracing JIT Compilation Technique}
\label{sec:meta-tracing_jit_compilation}

Typically, tracing \jit{} compilers record a representation of a program;
however, a meta-tracing \jit{} compiler traces the execution of an
\emph{interpreter} defined by a language builder. Meta-tracing \jit{}
compilation is just tracing compilation: in the sense that it compiles a
\emph{path} of a base-program, even if it has conditional branches.  If it has,
the compiled code will contain \emph{guards}, each of which is a conditional
branch to the interpreter execution from  that point.

\sloppy
\rpython{}~\cite{Bolz2009,BOLZ2015408}, a statically typed subset of Python
programming language, is a tool-chain for creating a high-performance \vm{}
with a trace-based \jit{} compiler. It requires a language builder for
implementing a bytecode compiler and interpreter definition for the
bytecode. Our prototype \baccaml{} is based on \rpython{}'s architecture. Before
describing the details of \baccaml{}, let us give an overview of \rpython{}'s
meta-tracing \jit{} compilation.

To leverage the RPython's \jit{} compiler, an interpreter developer should
annotate to help identify the loops in the \emph{base-program} that is going to
be interpreted. Figure~\ref{code:interp_rpython} shows an example of an interpreter
defined by a language developer. The example uses two annotations,
\emph{jit\_merge\_point} and \emph{can\_enter\_jit}. \emph{jit\_merge\_point}
should be put at the top of a dispatch loop to identify which part is the
main loop, and \emph{can\_enter\_jit} should be placed at the point where a
back-edge instruction can occur (where meta-tracing compilation might start).

\begin{figure}[t]
  \begin{minted}[fontsize=\scriptsize]{python}
def interp(bytecode):
  stack = []; sp = 0; pc = 0
  while True:
    jit_merge_point(reds=['stack','sp'],
                    greens=['bytecode','pc'])
    inst = bytecode[pc]
    if inst == ADD:
      v2, sp = pop(stack, sp)
      v1, sp = pop(stack, sp)
      sp = push(stack, sp, v1 + v2)
    elif inst == JUMP_IF:
      pc += 1; addr = bytecode[pc]
      if addr < pc: # backward jump
        can_enter_jit(reds=['stack','sp'],
                      greens=['bytecode','pc'])
      pc = addr
  \end{minted}
  \caption{Example interpreter definition written in RPython.}
  \label{code:interp_rpython}
\end{figure}

\begin{algorithm}[t]
  \caption{JitMetaTracing(rep, states)}\label{algo:meta_tracing}
  \SetKwInOut{Input}{input}
  \SetKwInOut{Output}{output}
  \SetKwData{Res}{residue}\SetKwData{Op}{op}\SetKwData{States}{states}\SetKwData{Rep}{rep}
  \Input{Representations of a interpreter itself}
  \Input{States (e.g., virtual registers and memories) of an interpreter itself}
  \Output{The resulting trace of the hot spot in a base-program}
  \SetKwData{Enstates}{entry\_states}
  \Enstates $\leftarrow$ \States\;
  \Repeat{\Op != jit\_merge\_point $\land$ \Enstates != \States}{
    \Res $\leftarrow$ [ ]
    \tcp*[l]{Data to store the result}
    \Op $\leftarrow$ \Rep.current\_operation(\States)\;
    \uIf{\Op = conditional branch}{
      \If{op has red variables}{
        guard $\leftarrow$ \Op.mk\_guard(\States)\;
        \Res.append(guard)\;
      }
      eval(\Op, \States, \Res)\;
    }
    \uElseIf{\Op = function call to f}{
      inline $f$\;
    }
    \Else{
      eval(\Op, \States, \Res)\;
    }
  }
  \Return \Res\;
\end{algorithm}

\begin{algorithm}[t]
  \caption{Eval(op, states, residue)}\label{algo:meta_tracing_eval}
  \SetKwInOut{Input}{input}
  \SetKwData{Residue}{residue}\SetKwData{Op}{op}\SetKwData{States}{states}
  \uIf{\Op has red variable}{
    \Op.const\_fold(\States)\;
    \Residue.append(\Op)\;
  }
  \Else{
    \Op.execute(\States)\;
  }
\end{algorithm}

Algorithms~\ref{algo:meta_tracing} and~\ref{algo:meta_tracing_eval} illustrate the
meta-tracing \jit{} compilation algorithm in pseudocode.
The procedure \textit{JitMetaTracing} takes the following arguments:
\textit{rep} -- a representation for the interpreter and \textit{states} --
the state of the interpreter just in starting to trace.
A meta-tracing \jit{} compiler records the execution and checks the operands in
the executed operations. It uses \textit{red} and \textit{green} colors for
recognizing runtime information. The color \textit{red} means ``a variable in a
base language''; hence, red variables are
used for calculating the result of a base-program. The color \textit{green}
indicates ``a variable in an interpreter'', then the compiler will optimize
this variable by constant-folding or
inlining. If all the operands in one operation are green, the operation is only
used for calculation in an interpreter, and therefore the compiler executes
it. If at least one variable is red, the compiler recognizes the operation is in
a base-program and writes to the \textit{residue}.


One significant advantage of a meta-tracing \jit{} compilation strategy
depicted by \rpython{} is that it is more comfortable to write interpreters in
comparison to the abstract-syntax-tree (\astree{}) rewriting specialization in
Truffle. In~\cite{Marr:2015:TVP:2814270.2814275}, Marr and Ducasse stated
that a significant difference between \rpython{} and \truffle{} is the number of
optimizations a language implementer needs to apply in order to reach the same
level of performance. In their experiments,
SOM~\cite{Haupt:2010:SFV:1822090.1822098} built with \rpython{} achieves
excellent performance without adding many optimizations. On the other hand, SOM
built with \truffle{} without any additional optimizations performs one order of
magnitude worse than the meta-tracing. By adding some optimizations, SOM with
\truffle{} reaches the same level as that of SOM with \rpython{}. According to
the result, they concluded that the meta-tracing strategy has significant
benefits from an engineering perspective~\cite{Marr:2015:TVP:2814270.2814275}.

\section{Approach}
\label{sec:architecture_overview}

\sloppy
This section explains the trade-offs between trace- and method-based compilation and
introduces our approach to solve the problem.

\subsection{Trade-Offs Between Trace-Based Compilation and Method-Based Compilation
  Strategies}
\label{sec:traceoffs_jit}

The advantages of the method-based compilation strategy are the following: First,
the same optimization techniques used in \aot{} can be applied for method-based
compilation. Thus, it can leverage existing \aot{} compiler engines such as
GCC~\cite{gcc} and LLVM~\cite{llvm}. Second, a compilation unit has the
complete control flow of a target method.
Therefore, method-based compilation can be applied not only for various types of
programs and also for programs that are not suitable for trace-based
compilation~\cite{Inoue2011,Marr:2015:TVP:2814270.2814275}.

\sloppy
However, when a method-based compiler compiles a method with many biased branches,
the compiled code includes colds spots of the method. This makes compilation time
longer than when applying trace-based compilation strategy. Furthermore, it requires
well-planned inlining to a method for reducing the overhead of a function call. When
it applies aggressive inlining to a method with deeply-nested function calls, the
compiled code's size increases, leading to longer compilation time.

The trace-based compilation strategy, however, can apply many optimization
techniques~\cite{Bolz:2011:RFM:2069172.2069181}, including constant-subexpression
elimination, dead-code elimination, constant-folding, and register allocation
removal~\cite{Bolz:2011:ARP:1929501.1929508}, since compilation code represents
only one execution path. Thus, this strategy gets better results with specific
programs with branching possibilities or
loops~\cite{Bauman:2015:PTJ:2784731.2784740,Huang2016}. Moreover, it can execute
aggressive function inlining at low cost, since a trace-based \jit{} compiler
tracks and records the execution of a program so that a resulting trace will include
an inlined function call~\cite{Gal:2006:10.1145/1134760.1134780}. This leads to
reducing overheads of a function call and creating chances for further
optimization. However, this strategy performs worse at programs with complex control
flow because of the mismatch between tracing and execution~\cite{Huang2016}. In
contrast, method-based compilation performs better in such programs.

\subsection{Meta-Hybrid JIT Compiler Framework}
\label{sec:meta_hybrid_jit_compiler_framework}


We propose a \emph{meta-hybrid \jit{} compiler framework} to overcome the trade-offs
explained above, and the prototype implementation namely \baccaml{}.
The framework is a \emph{meta}-\jit{} compiler framework; therefore, a language
developer can generate a VM with a hybrid \jit{} compiler by writing interpreter
definition. A generated \jit{} compiler is a hybrid of the trace- and method-based
compilations as it can select an execution path or a function as a compilation
unit. The compiled code from the two types of strategies can work together in a
single execution.

\begin{figure}[t]
  \centering
  \begin{subfigure}[t]{1.0\columnwidth}
    \centering
    \includegraphics[width=0.75\columnwidth]{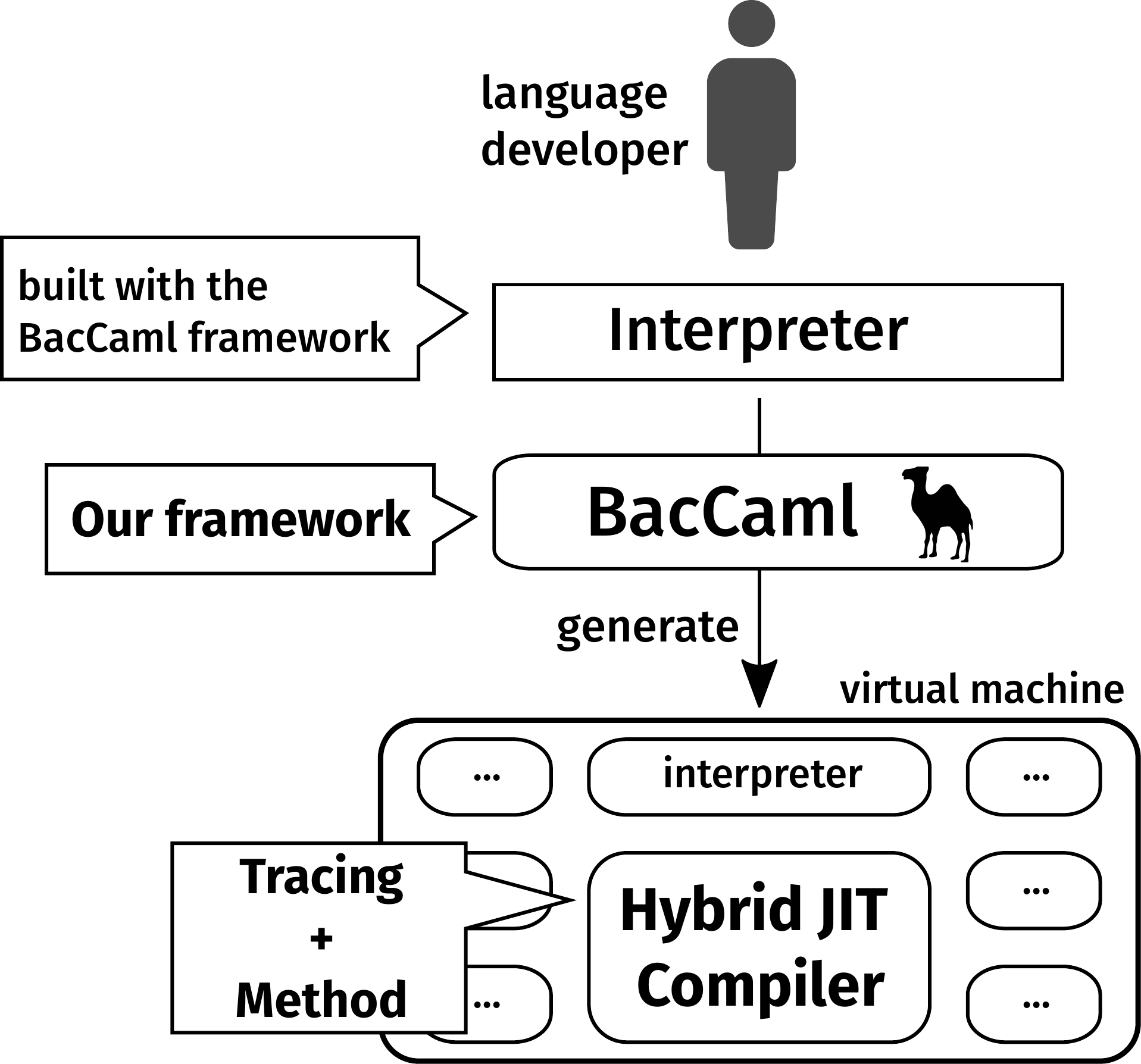}
    \caption{A virtual machine generation with our meta-hybrid JIT compiler
      framework.}
    \label{fig:metajit_vmgen}
  \end{subfigure}
  \begin{subfigure}[t]{1.0\columnwidth}
    \centering
    \includegraphics[width=0.9\columnwidth]{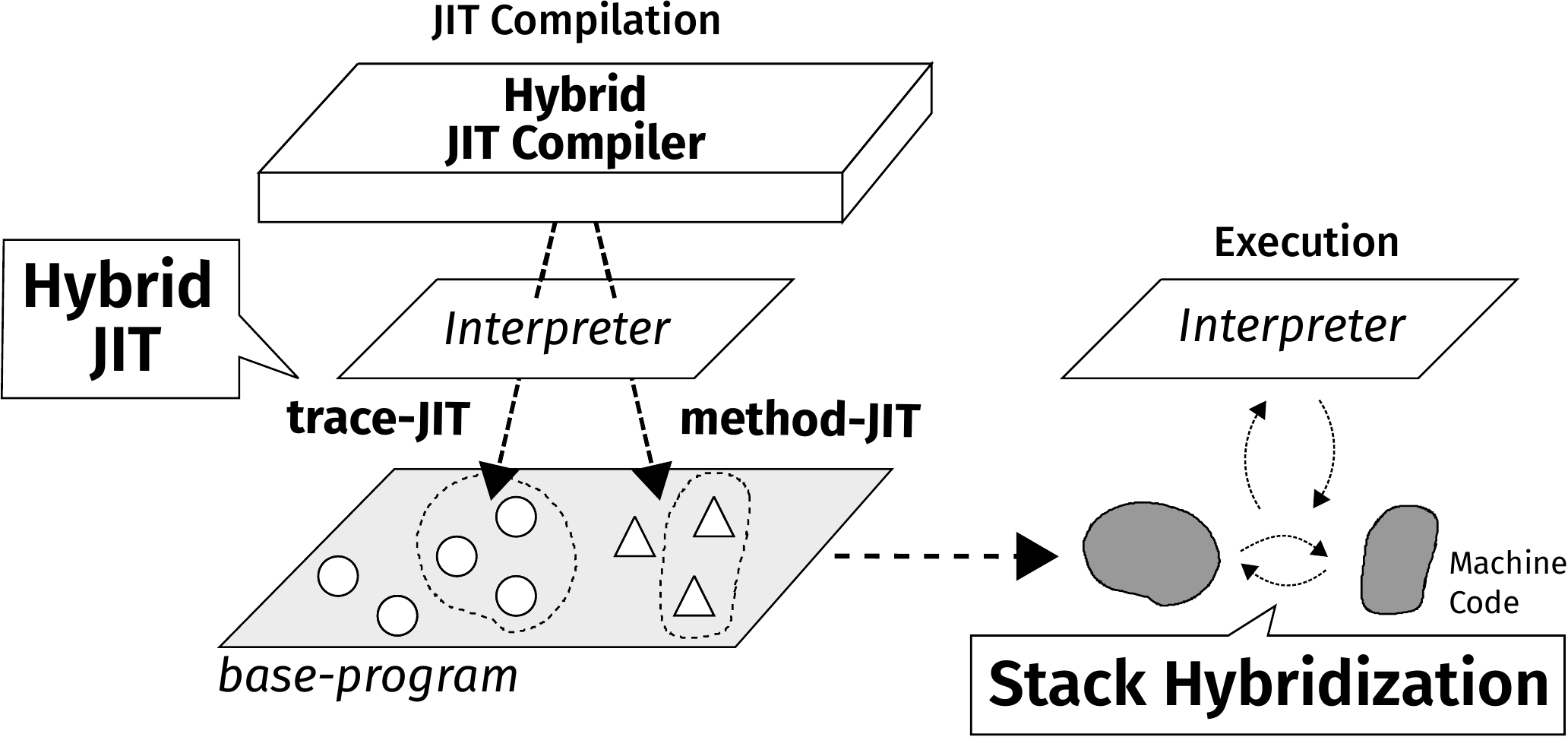}
    \caption{A runtime overview of a generated hybrid \jit{} compiler.}
    \label{fig:metajit_comp_runtime}
  \end{subfigure}
  \caption{The overviews of our meta-hybrid \jit{} compiler framework.}
  \label{fig:metajit_overview}
\end{figure}

The basic idea of achieving hybrid compilation is realizing method-based compilation
by extending a (meta-) tracing compiler and mixing them.
Since a trace has no control flow and inlines a function, we create our
method-based compilation by customizing the tracing \jit{} compilers' features to
cover all the paths in a method. Therefore, our meta-hybrid \jit{} compiler framework
shares its implementations between the two trace- and method-based compilers. Details
are explained in Section~\ref{sec:method_jit_by_trace}.

In addition to the leverage strength of the two \jit{} compilation strategies, we aim
to resolve the path-divergence problem by selectively applying method-based
compilation to the functions that cause the problem trace compilation to the other
parts of a program. Since this proposal focuses on combining the different two
strategies, dynamically selecting a suitable strategy depending on target programs'
structure is left as future work.

Figure~\ref{fig:metajit_overview} gives an overview of our meta-hybrid \jit{}
compiler framework. As shown in Figure~\ref{fig:metajit_vmgen}, when a language
developer writes interpreter definition with the framework, the framework can
generate a VM with a hybrid \jit{} compiler. Figure~\ref{fig:metajit_comp_runtime}
overviews our hybrid compilation and runtime by our framework. At runtime, the
generated hybrid \jit{} compiler applies different strategies to different parts of
a base-program. Further, machine codes generated from different strategies can move
back and forth with each other by \emph{Stack Hybridization}, which is illustrated in
Section~\ref{sec:stack_hybridization}.

\section{Mixing The Two Compilation Strategies in Meta-Level}
\label{sec:mixing_the_two_compilationss}

In this section, we first describe how to construct method-based compilation based on
(meta-) trace-based compilation. We then explain how to cooperate with them in a meta
\jit{} compiler framework. In this work, we merely aim at achieving \emph{simple}
method compilation; i.e., advanced optimization techniques used in existing method
\jit{} compilers are left for future work.

\subsection{Method-Based Compilation by Tracing}
\label{sec:method_jit_by_trace}

To construct method-based compilation by utilizing a trace-based compilation, we have
to cover all paths of a function. In other words, we need to determine the \emph{true
  path} and decrease the number of guard failures that occur to solve the
path-divergence problem when applying trace-based compilation.

We propose \emph{method \jit{} by tracing} by customizing the following features of
trace-based compilation: (1) trace entry/exit points, (2) conditional branches (3)
loops, (4) function calls. In the following paragraphs, we explain in detail how to
``trace'' a method by modifying these features. Note that method \jit{} by tracing is
more naive than other state-of-the-art method-based \jit{} compilers, since this
method-based compilation is designed for applying for programs with complex control
flow, which causes performance degradation problem in a trace-based compilation, and
we apply a trace-based compilation for other programs. Thus, the trace-based compiler
is the primary compiler, and the method-based compiler is the secondary compiler in
our system.

\paragraph{Trace entry/exit points}

\begin{algorithm}[t]
  \caption{JitMetaMethod(rep, states, residue)}
  \label{algo:trace_method_entry_exit}
  \SetKwInOut{Input}{input}
  \SetKwInOut{Output}{output}

  \SetKwData{Res}{residue}\SetKwData{Op}{op}\SetKwData{States}{states}\SetKwData{Rep}{rep}
  \SetKwFunction{FTraceCond}{TraceCond}\SetKwFunction{FTraceLoop}{TraceLoop}
  \SetKwFunction{FTraceFun}{TraceFunction}

  \Input{Representations of an interpreter itself.}
  \Input{States (e.g., virtual registers and memories) of an interpreter itself.}
  \Input{An array data structure that records an executed instruction.}
  \Output{The trace of a target method in a base-program}

  \SetKwRepeat{Do}{do}{while}
  \eIf{op = method\_entry}{
    \Res $\leftarrow$ [ ]\;
    \Do{op != return}{
      \uIf{op = conditional branch}{
        \FTraceCond(\Rep, \States, \Res)\;
      }
      \uElseIf{op = loop entry}{
        \FTraceLoop(\Rep, \States, \Res)\;
      }
      \uElseIf{op = function call to f}{
        \FTraceFun(\Rep, \States, \Res)\;
      }
      \uElse{
        eval(\Op, \States, \Res)\;
      }
      \Op $\leftarrow$ \Rep.get\_next(\Op)\;
    }
    \Return{\Res}\;
  }{
    \Return{}\;
  }
\end{algorithm}

\begin{algorithm}[t]
  \caption{TraceCond(rep, states, residue)}
  \label{algo:trace_cond_branch}
  \SetKwData{Regs}{regs}\SetKwData{Mems}{mems}
  \SetKwData{Res}{residue}\SetKwData{Op}{op}\SetKwData{States}{states}\SetKwData{Rep}{rep}
  \SetKwData{TrThen}{trace\_then}\SetKwData{TrElse}{trace\_else}\SetKwData{TrIf}{trace\_ifexp}
  \SetKwRepeat{Do}{do}{while}
  \Regs, \Mems $\leftarrow$ [ ], [ ]\;
  \Do{\Op~!=~return}{
    \Regs.store(\States.get\_reg())\;
    \Mems.store(\States.get\_mem())\;
    \TrThen $\leftarrow$ \textsc{JitMetaMethod}(\States)\;
    \States.restore(\Regs, \Mems)\;
    \TrElse $\leftarrow$ \textsc{JitMetaMethod}(\States)\;
    \tcp{construct if exp including trace\_then and trace\_else}
    \TrIf $\leftarrow$ \Begin{
      \eIf{\Op.const\_fold(\States)}{
        \TrThen\;
      }{
        \TrElse
      }}
    \Res.append(\TrIf)\;
    \Op $\leftarrow$ \Rep.next\_of(\Op)\;
  }
\end{algorithm}

\begin{figure}[t]
  \footnotesize
  \centering
  \begin{subfigure}[t]{.55\linewidth}
    \centering
    \begin{tikzpicture}[
  scale=0.5,
  node distance=.75cm,
  normal/.style={draw,circle,fill=white,text width=1.5mm}]

  \node[normal,label=above:{\small entry}, very thick] (A) {A};
  \node[normal,below of=A] (B) {B};
  \node[normal,below left of=B] (C) {C};
  \node[normal,below right of=B] (D) {D};
  \node[normal,below of=C] (E) {E};
  \node[normal,below of=D] (F) {F};
  \node[normal,below right of=E] (G) {G};
  \node[normal,below of=G, very thick, label=below:{\small return}] (H) {H};

  \path[-stealth] (A) edge (B);
  \draw[-stealth] (B) edge (C);
  \draw[-stealth] (B) edge (D);
  \path[-stealth] (C) edge (E);
  \path[-stealth] (E) edge (G);
  \path[-stealth] (D) edge (F);
  \path[-stealth] (F) edge (G);
  \path[-stealth] (G) edge (H);

\end{tikzpicture}
\quad
\begin{tikzpicture}[
  scale=0.5,
  node distance=.75cm,
  normal/.style={draw,circle,fill=white,text width=1.5mm},
  background/.style={rectangle, rounded corners, fill=gray!10, draw=black!20, inner sep=0.4mm}]

  \node[normal, label=above:{\small entry}, very thick] (A) {A};
  \node[normal, below of=A] (B) {B};
  \node[normal, below left of=B] (C) {C};
  \node[normal, below right of=B] (D) {D};
  \node[normal, below of=C] (E) {E};
  \node[normal, below of=D] (F) {F};
  \node[normal, below of=E] (G1) {G};
  \node[normal, below of=F] (G2) {G'};
  \node[normal, below of=G1, very thick, label=below:{\small return}] (H1) {H};
  \node[normal, below of=G2, very thick, label=below:{\small return}] (H2) {H'};

  \path[-stealth]
    (A) edge (B)
    (B) edge (C)
    (B) edge (D)
    (C) edge (E)
    (E) edge (G1)
    (D) edge (F)
    (F) edge (G2)
    (G1) edge (H1)
    (G2) edge (H2);

  \begin{scope}[on background layer]
    \node [background,fit=(A) (H1) (H2),label={80:{\scriptsize \textcolor{gray}{tr.1}}}] {};
  \end{scope}

\end{tikzpicture}
    \caption{Handling of a conditional branch. In this program, A
      represents a method entry, and B -- C -- D represents a conditional
      branch.}
    \label{fig:mj_if_graph}
  \end{subfigure}
  \hfill
  \begin{subfigure}[t]{.4\linewidth}
    \centering
    \begin{tikzpicture}[
  node distance=.75cm,
  background/.style={rectangle, rounded corners, fill=gray!10, draw=black!20,
    inner sep=0.5mm},
  node/.style={circle, draw, fill=white, text width=1.5mm}]

  \node[node,very thick,label=above:{\small entry}] (A) {A};
  \node[node,below of=A] (B) {B};
  \node[node,below of=B] (C) {C};
  \node[node,below of=C] (D) {D};
  \node[node,below of=D] (E) {E};
  \node[node,below of=E,very thick,label=below:{\small return}] (F) {F};

  \draw[-stealth] (A) -- (B);
  \draw[-stealth] (B) -- (C);
  \draw[-stealth] (C) -- (D);
  \draw[-stealth] (D) -- (E);
  \path[-stealth] (D.east) edge [bend right] (B.south east);
  \draw[-stealth] (E) -- (F);
\end{tikzpicture}
\quad
\begin{tikzpicture}[
  node distance=.77cm,
  background/.style={rectangle, rounded corners, fill=gray!10, draw=black!20,
    inner sep=0.5mm},
  node/.style={circle, draw, fill=white, text width=1.5mm}
  ]
  \tikzstyle{nn} = [draw,circle,fill=white]
  \tikzstyle{background} = [rectangle, rounded corners, fill=gray!10,
  draw=black!20]

  \node[nn,very thick,label=above:{\small entry}] (A) {A};
  \node[nn,below of=A,] (B) {B};
  \node[nn,below of=B,] (C) {C};
  \node[nn,below of=C] (D) {D};
  \node[nn,below right of=D,node distance=1cm] (E) {E};
  \node[nn,below of=E,very thick,label=below:{\small return}] (F) {F};
  \draw[-stealth] (A) -- (B);
  \draw[-stealth] (B) -- (C);
  \draw[-stealth] (C) -- (D);
  \draw[-stealth] (D) -- (E);
  \path[-stealth] (D.east) edge [bend right] (B.south east);
  \draw[-stealth] (E) -- (F);

  \begin{scope}[on background layer]
    \node [background,fit=(A),label={20:{\scriptsize
        \textcolor{gray}{tr.1}}}] {};
    \node [background,fit=(B)(C)(D),label={65:{\scriptsize
        \textcolor{gray}{tr.2}}}] {};
    \node [background,fit=(E)(F),label={90:{\scriptsize
        \textcolor{gray}{tr.3}}}] {};
  \end{scope}
\end{tikzpicture}
    \caption{Handling of a loop. In this program, B -- C -- D represent a loop,
      and E -- F is a successor of the loop B -- C -- D.}\label{fig:mj_loop_graph}
  \end{subfigure}

  \begin{subfigure}[t]{1.0\linewidth}
    \centering
    \begin{tikzpicture}[
  scale=0.8,
  background/.style={rectangle, rounded corners, fill=gray!10, draw=black!20,
    inner sep=0.4mm},
  transparent/.style={draw,draw opacity=0.3,fill opacity=0.3},
  nn/.style={draw,circle,fill=white,node distance=.9cm}]

  \node[nn,very thick,label=north:{\scriptsize entry}] (A) {A};
  \node[nn,below of=A] (B) {B};
  \node[nn,below of=B] (C) {C};
  \node[nn,very thick,below of=C,label=south:{\scriptsize return}] (D) {D};
  \node[nn,above right of=C,node distance=1.2cm] (E) {E};
  \node[nn,below right of=C,node distance=1.2cm] (F) {F};

  \draw[-stealth] (A) -- (B);
  \draw[-stealth] (B) -- (C);
  \draw[-stealth,densely dotted] (E) -- (F);
  \draw[-stealth] (C) -- node[midway,above,sloped] {\scriptsize call} (E);
  \draw[-stealth] (F) -- node[midway,above,sloped] {\scriptsize return} (C);
  \draw[-stealth] (C) -- (D);

  \begin{scope}[on background layer]
    \node[draw,rectangle,fit=(E)(F),label=above:{\scriptsize other fun.}] {};
  \end{scope}

\end{tikzpicture}
\quad \quad
\begin{tikzpicture}[
  scale=0.8,
  node distance=.9cm,
  background/.style={rectangle,rounded corners,fill=gray!10,draw=black!20,inner
    sep=0.4mm},
  transparent/.style={draw,draw opacity=0.3,fill opacity=0.3},
  nn/.style={draw,circle,fill=white,node distance=.9cm}]

  \node[nn,very thick,label=above:{\scriptsize entry}] (A) {A};
  \node[nn,below of=A] (B) {B};
  \node[nn,below of=B] (C) {C};
  \node[nn,very thick,below of=C,label=below:{\scriptsize return}] (D) {D};
  \node[nn,above right of=C,node distance=1.2cm,transparent] (E) {E};
  \node[nn,below right of=C,node distance=1.2cm,transparent] (F) {F};

  \draw[-stealth] (A) -- (B);
  \draw[-stealth] (B) -- (C);
  \draw[-stealth,transparent,densely dotted] (E) -- (F);
  \draw[densely dotted, -stealth] (C) -- node[midway,above,sloped] {\scriptsize call} (E);
  \draw[-stealth,transparent] (F) -- node[midway,above,sloped] {\scriptsize return} (C);
  \draw[-stealth] (C) -- (D);

  \begin{scope}[on background layer]
    \node [background,fit=(A)(D),label={80:{\scriptsize \textcolor{gray}{tr.1}}}] {};
    \node [rectangle,transparent,solid,fit=(E)(F),label=above:{\scriptsize other fun.}] {};
  \end{scope}

\end{tikzpicture}
    \caption{Handling of a function call. A -- B -- C -- D and E -- F are
      functions. In this program, (A -- B -- C -- D) calls (E -- F) at C. Note
      that only target function (A -- B -- C --D) is compiled.}
    \label{fig:mj_call_graph}
  \end{subfigure}
  \caption{Examples how our method-based compilation works. Each left-hand side
    is the control-flow of a target base-program that represents one method, and each
    right-hand side is a result. ``entry'' and ``return'' means the entry point and
    exit poitn of a target method, respectively.}
\end{figure}
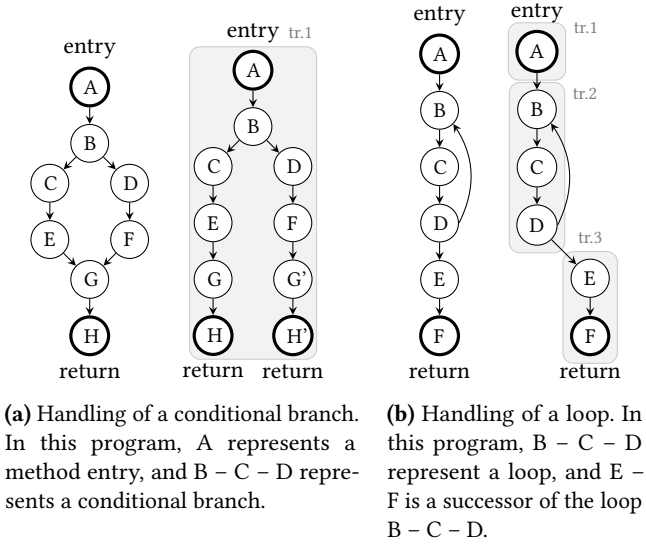
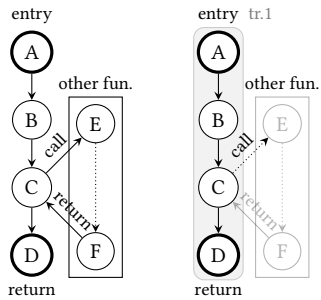

Trace-based \jit{} compilers~\cite{Gal:2006:10.1145/1134760.1134780,Gal2009}
generally compile loops in the base-program; therefore, they start to trace at
the top of a loop and end when the execution returns to the entry point. To
assemble the entire body of a function, we modify this behavior to trace from
the top of a method body until a \verb|return| instruction is reached (see
Algorithm~\ref{algo:trace_method_entry_exit}).

\paragraph{Conditional branches}

\sloppy
When handling a conditional branch, trace-based \jit{} compilers convert a
conditional branch into a guard instruction and collects instructions that are
executed. When execution method-based compilation, however, we must compile both
sides of conditional branches. To achieve this, a tracer that records executed
instructions must return to the branch point and restart tracing the other side as
well. As shown in Algorithm~\ref{algo:trace_cond_branch}, the tracer in our
constructed method-based \jit{} compiler has to trace both then and else sides so
that it \emph{backtracks} to the beginning of a conditional branch when it reaches
the end of one side and continues to trace the other side. Before starting to trace
one side, the tracer stores its states (e.g., the data stored in the tracer's virtual
registers and memories) in already prepared arrays. For just backtracking, the tracer
\emph{restores} those states and continues to the other side.

Figure~\ref{fig:mj_if_graph} shows an example describing how the tracer for
method-based compilation works. On the left side, node A is the method entry, nodes
B -- C -- D form a conditional branch, and node E is the end of
this method. The tracer starts to trace at A. On reaching a conditional branch (B),
the tracer then stores its state and follows one side (B -- C -- E -- G -- H). On
reaching \emph{return} instruction (H), the tracer finally backtracks to B and
resumes to trace the other side (B -- D -- F -- G -- H) by restoring the already
saved data.

There is a risk of an exponential blow-up of compiled code when tracing a
program that has many nested conditionals. To avoid generating too big native
code, when the compiler detects too many branches in a target program part, our
system stops the method-based compiler to trace. Instead, our system switches to
apply trace-based compilation for such a program.

\paragraph{Loops}

\begin{algorithm}[t]
  \caption{TraceLoop(rep, states, residue)}
  \label{algo:mj_loop}
  \SetKwData{Res}{residue}
  \SetKwData{Op}{op}
  \SetKwData{State}{state}
  \SetKwData{States}{states}
  \SetKwData{Rep}{rep}
  \SetKwData{Enstate}{entry\_state}
  \SetKwRepeat{Do}{do}{while}

  \SetKwData{Res}{residue}
  \SetKwData{ResAfter}{residue\_loop\_after}

  \Op $\leftarrow$ \Rep.get\_op()\;
  \Do{\Op != return}{
    \uIf{\Op = back-edge to the entry}{
      \Res.append(\textbf{jump} to entry)\;
    }
    \uElseIf{\Op = loop-exit}{
      loop\_after\_state $\leftarrow$ \Rep.next\_of(\Op).get\_states()\;
      loop\_after\_trace $\leftarrow$ \textsc{JitMetaMethod}(\Rep,
      loop\_after\_state)\;
      \Res.append(\textbf{jump} to loop\_after\_trace)\;
      \ResAfter.append(loop\_after\_trace)\;
    }
    \uElse{
      eval(\Op, \States, \Res)\;
    }
    \Op $\leftarrow$ \Rep.next\_of(\Op)\;
  }
  \Res.append(\ResAfter)
\end{algorithm}

Our method-based compiler does not handle loops specially. While a trace-based
compiler compiles loops as a straight-line path, our method-based compiler
compiles not only the body a target loop, but also the successors of it.

Algorithm~\ref{algo:mj_loop} illustrates how the tracer for method-based compilation
traces a loop. When the tracer finds the entry point, it starts to analyze the body
of a function to find a back-edge and loop-exit instruction. When the tracer traces a
back-edge, as with a trace-based compilation, leaves an instruction to \verb|jump| to
the entry. When the tracer leaves a loop-exit instruction, it also traces the
destination of a loop-exit instruction and leaves a \verb|jump| instruction to go to
the outside of this loop.

Figure~\ref{fig:mj_loop_graph} shows an example of how to handle a loop. In this
example, our method-based compiler compiles a single loop into three trace parts. The
first one (tr.1) is up to the loop entry, the second one is the loop itself (tr.2) of
the loop, and the third one is the successor of the loop (tr.3).

\paragraph{Function calls}
\label{par:func_call}

\begin{algorithm}[t]
  \caption{TraceFunction(rep, states, residue)}
  \SetKwData{Res}{residue}
  \SetKwData{Op}{op}
  \SetKwData{States}{states}
  \SetKwData{Rep}{rep}
  \label{algo:trace_funcall}
  \SetKwRepeat{Do}{do}{while}
  \Do{op != return}{
    \If{op = function call to f}{
      \Res.append(\textit{call to f}) \tcp*[l]{not following but leaving the
        instruction ``call f''}
      \tcp{continue to trace successors}
    }
  }
\end{algorithm}

Whereas a trace-based \jit{} compiler will inline function calls, our method-based
\jit{} compiler will not inline, but emit a call instruction code and continue
tracing. We don't inline function because our method-based compilation is designed to
apply only for programs with the path-divergence problem. If a target program needs
inlining, we will apply trace-based compilation for it, since trace-based compilation
can automatically perform function inlining. Thus, our method-based compilation is so
naive that it is not equivalent to other method-based \jit{} compilers.

To remain a function call in a resulting trace, we have to inform which part is
represented to a base-program function call in an interpreter definition. Therefore,
we need to implement the specific interpreter style shown in the left-hand side of
Figure~\ref{code:user_host_stack_style} (we call this style \emph{host-stack style}
here). By writing in host-stack style, the tracer can detect which part is a
base-program's method invocation and leave a call instruction in a resulting
trace. Figure~\ref{fig:mj_call_graph} shows how the tracer compiles a function
call. In this example, the tracer eventually generates one trace, including a call
instruction (tr.1).

In trace-based compilation, however, a meta-tracing \jit{} compiler can work
efficiently in a specific way as shown in the right-hand side of
Figure~\ref{code:user_host_stack_style} (we call this \emph{user-stack style} here).

In the next section, we organize why we need the two different two stack styles.

\begin{figure}[t]\centering
\begin{subfigure}[b]{.48\columnwidth}
\begin{minted}{python}
if opcode == CALL:
  addr = self.bytecode[pc]
  # call the `interp'
  # recursively
  res = self.interp(addr)
  user_stack.push(res)
  pc += 1
elif opcode == RETURN:
 # return a top of
 # `user-stack'
  return user_stack.pop()
\end{minted}
\caption{\sloppy Example interpreter written in host-stack style.}
\label{code:host_stack_style}
\end{subfigure}
\begin{subfigure}[b]{.48\columnwidth}
\begin{minted}{python}
if opcode == CALL:
  addr = self.bytecode[pc]
  pc += 1
  # push a return addr to
  # `user-stack'
  ret_addr = W_IntObject(pc)
  user_stack.push(ret_addr)
  if addr < pc:
    can_enter_jit(..)
  # jump to a callee function
  pc = t
elif opcode == RETURN:
  v = user_stack.pop()
  # restore already pushed
  # return addr
  addr = user_stack.pop()
  user_stack.push(v)
  if addr < pc:
    can_enter_jit(..)
  # jump back to the caller
  # function
  pc = addr
\end{minted}
\caption{\sloppy Example interpreter written in user-stack style.}
\label{code:user_stack_style}
\end{subfigure}
\caption{Interpreter definition styles. For managing a return
   address/value, left-hand side style uses a host-language's (system
   provided) stack, but right-hand side uses a developer-prepared stack data
   structurpe.}
 \label{code:user_host_stack_style}
\end{figure}

\subsection{The Two Stack Styles}
\label{sec:two_stack_styles}

A compiler implemented by a meta-tracing JIT compiler framework has two options to
represent a call-stack, namely the host-stack and the user-stack. Those two options
are chosen based on the way of implementing function call/return operation in the
interpreter. The host-stack style interpreter, which can be found in
PyPy~\cite{PyPyInterp} and Topaz~\cite{TopazInterp}, uses the host-language's
function call/return for the base-language's call/return. The user-stack style
interpreter, as in Pycket~\cite{PycketInterp} and Pyrlang~\cite{PyrlangInterp},
manages return addresses of the base-language's function calls in a user-defined data
structure in the interpreter.

Each of the two compilation strategies requires one of those two stack
styles. Concretely, the method-based compilation requires the host-stack style,
whereas the trace-based compilation prefers the user-stack style. While this causes
the combination problem explained in the next section, we describe the relationship
between the compilation strategies and the stack styles after explaining the
differences between them.

\paragraph{Method-based compilation requires the host-stack \\ style.} The method-based
compilation requires to confine the compilation target to one
function/method~\footnote{Even with function inlining, and the compiler must stop
  inlining at some point.} in a base-program, which means the compiler needs to leave
a function call operation in the base program as a function call instruction in the
compiled code and continue compilation of the subsequent operations in the caller
function.

With a host-stack style interpreter, since the function call operation is
implemented as a call to the interpreter function (i.e.,
\emph{res = self.interp(addr)}) in Figure~\ref{code:host_stack_style}), it is easy to
leave the function call as it is and to continue compiling the subsequent
operations.

With a user-stack style interpreter, however, the function call operation is
implemented by manipulating the interpreter's program counter (i.e.,
\emph{user\_stack.push(ret\_addr); pc = t} in Figure~\ref{code:user_stack_style}). To
compile this into a function call machine instruction and continue compilation of the
subsequent operations on the caller's side, the compiler needs to trace the
interpreter with the state after returning from the callee function without running
the callee function. It is not easy, if not impossible.

\begin{figure}[t]
  \centering
  \begin{multicols*}{2}
  \begin{minted}[fontsize=\scriptsize]{python}
    # Example sum function
    def sum(n):
      if n <= 1: return 1
      else: return n + sum(n-1)

    sum(10000)
  \end{minted}
  \hrule
  \begin{minted}[fontsize=\tiny]{python}
    # Resulting trace from `sum'
    dbg_mrg_point(0,0,#0 LOAD_FAST')
    dbg_mrg_point(0,0,#3 LOAD_CONST')
    dbg_mrg_point(0,0,#6 COMPARE_OP')
    dbg_mrg_point(0,0,#9 POP_JMP_IF_FLS')
    dbg_mrg_point(0,0,#16 LOAD_FAST')
    dbg_mrg_point(0,0,#19 LOAD_GLOBAL')
    dbg_mrg_point(0,0,#22 LOAD_FAST')
    dbg_mrg_point(0,0,#25 LOAD_CONST')
    ...
    dbg_mrg_point(5,5,#33 RETURN_VALUE')
    dbg_mrg_point(4,4,#32 BINARY_ADD')
    dbg_mrg_point(4,4,#33 RETURN_VALUE')
    dbg_mrg_point(3,3,#32 BINARY_ADD')
    dbg_mrg_point(3,3,#33 RETURN_VALUE')
    dbg_mrg_point(2,2,#32 BINARY_ADD')
    dbg_mrg_point(2,2,#33 RETURN_VALUE')
    dbg_mrg_point(1,1,#32 BINARY_ADD')
    dbg_mrg_point(1,1,#33 RETURN_VALUE')
    dbg_mrg_point(0,0,#32 BINARY_ADD')
    dbg_mrg_point(0,0,#33 RETURN_VALUE')
  \end{minted}
  \end{multicols*}
  \caption{Example non-tail recursive function and its compiled trace in PyPy.}
  \label{code:sum-and-resulting-trace}
\end{figure}

\paragraph{Trace-based compilation prefers the user-stack style.} The trace-based
compilation prefers a user-stack style interpreter to successfully compile programs
that heavily use recursive calls and to support languages that provide operators to
manipulate call stacks (e.g., first-class continuations).

When a base-program with a (non-tail) recursive call (e.g., \emph{sum(n)} in
Figure~\ref{code:sum-and-resulting-trace}) runs, it roughly performs a repetition of
function calls followed by a repetition of function returns. A tracing compiler can
compile the traces into two loops with a user-stack style interpreter, respectively
correspond to the call/return repetitions. This is because the user-stack style
interpreter realizes the base function call/return as jumps.

However, with a host-stack style interpreter, a tracing compiler can either
terminates compilation upon the function call operation in the interpreter or inline
the call operation. The former will yield very short compiled code fragments, which
entails a significant amount of overheads. The latter can work well when nested
levels of calls are not deep. When the nested levels get more in-depth, it can lead
to code bloat. Figure 5 shows the compilation of a non-tail recursive function in
PyPy, which uses the host-stack style interpreter. The sequence below the horizontal
bar is the intermediate compiled code, where we can see that the compiler performs
function inlining. For functions with a larger body, this approach will cause code
bloating or limited levels of inlining.

\subsection{Combination Problem}
\label{sec:integration_problem}

\begin{figure}[t]
  \centering
  \begin{subfigure}[b]{.8\linewidth}
    \centering
    \includegraphics[width=\columnwidth]{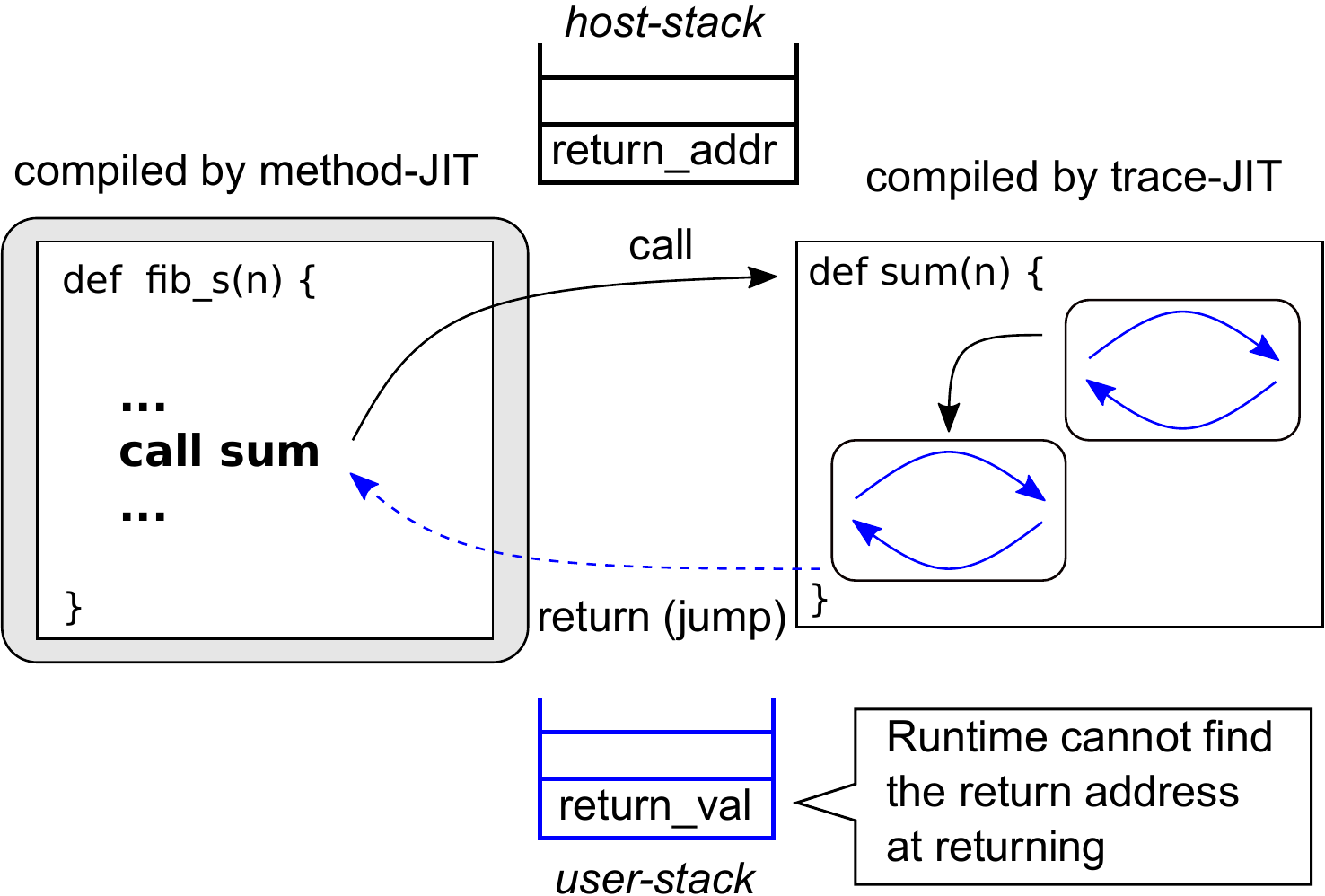}
    \caption{Calling a trace-compiled code from a
      method-compiled function.}
    \label{fig:combination_problem_1}
  \end{subfigure}
  \begin{subfigure}[b]{0.8\linewidth}
    \centering
    \includegraphics[width=\columnwidth]{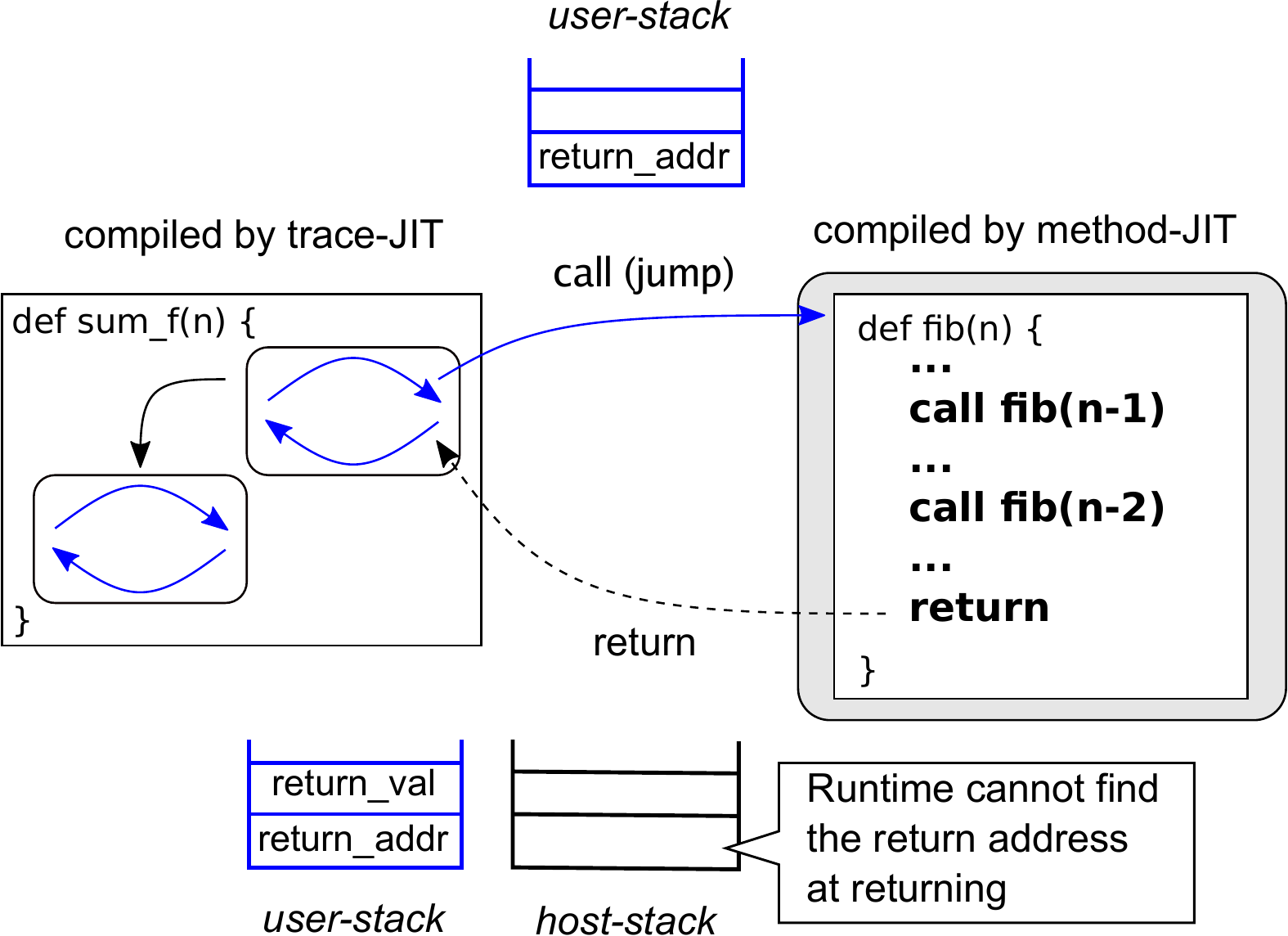}
    \caption{Calling a method-compiled function from a
      trace-compiled code.}
    \label{fig:combination_problem_2}
  \end{subfigure}
  \caption{Example of Combination Problem. Gray background code is compiled by
    method \jit{}, and blue lined code is compiled by tracing \jit{}.}
  \label{fig:combination_problem}
\end{figure}

The reason why we cannot naively combine them is the following: the two compilations
require different interpreter implementation styles in function calls. Trace-based
compilation requires the \emph{user-stack style}, while method-based compilation
requires the \emph{host-stack style}. In other words, different types of compilations
use different stack frames for optimizing function calls. Because of this gap, the
runtime cannot call back and forth between native codes generated from the two
compilations. Trace-based compilation inlines a function call; therefore, there is no
function call instruction in the resulting trace. Whereas method-based compilation
``leaves'' a function call instruction in the resulting trace. We explain this
problem by using
Figure~\ref{fig:combination_problem}. Figure~\ref{fig:combination_problem_1} shows an
example that a method-compiled function calls a trace-compiled function, and
Figure~\ref{fig:combination_problem_2} shows an example that a method-compiled
function calls a trace-compiled function.

In the case that \verb|fib_s| (compiled by method-based compilation) calls
\verb|sum| (compiled by trace-based compilation) as shown in
Figure~\ref{fig:combination_problem}, the runtime puts a return address in the
host-stack. In \verb|sum|, the return value and return address are stored in the
user-stack. On returning from \verb|sum|, since the semantics of return is
defined as shown in Figure~\ref{code:user_host_stack_style}, the runtime attempts
to find a return address from a user-stack. However, the return address is
stored in a host-stack, and the runtime cannot return to the correct place.

In the case shown in Figure~\ref{fig:combination_problem_2}, \verb|sum_f| (compiled
by trace-based compilation) calls \verb|fib| (compiled by method-based compilation),
however the runtime puts its return address in the user-stack. When runtime returns
from \verb|fib|, it then attempts to find the return address from the host-stack, but
it fails to find the address and results in runtime-error because the return
address is pushed to the user-stack.

\subsection{Stack Hybridization}
\label{sec:stack_hybridization}



To overcome this problem, we also present \emph{\sh{}}, a mechanism to bridge the
native codes generated from different strategies. \sh{} manages different kinds
of stack frames, and generates machine code that can be mutually executed in
trace-JIT and method-JIT contexts. To use \sh{}, a language developer needs to
write an interpreter in the specific way: (1) For executing a call
instruction in the base language, developers put a special flag to indicate
which stack frame is used in a self-prepared stack data structure. (2)  For
executing a return, they have to branch to the return instruction of the base
language corresponding to the call by checking the already pushed flag.

Roughly speaking, the interpreter handles the call and return operations in the
following ways:

\begin{itemize}
\item When it calls a function under the trace-based compilation, it uses the
  user-stack; i.e., it saves the context information in the stack data
  structure, and iterates the interpreter loop. Additionally,  it leaves a flag
  ``user-stack'' in the user-stack.

\item When it calls a function under the method-based compilation, it uses the
  host-stack; i.e., it calls the interpreter function in the host language.
  Additionally, it leaves a flag ``host-stack'' in the user-stack.

\item When it returns from a function, it first checks a flag in the
  user-stack. If the flag is ``user-stack'', it restores the context information
  from the user-stack.
  Otherwise, it returns from the interpreter function  using the host-stack.
\end{itemize}

To support behaviors, we introduce an interpreter implementation style, which enables
to embed both styles into a single interpreter and switch its behavior depending on
the flag. Figure~\ref{code:stack_hybridization_meta_interp} shows a sketch of special
syntax to support \sh{}. The important syntaxes are \verb|is_mj| pseudo
function, and \verb|US/HS| special flags. \verb|US| means a flag ``user-stack'', and
\verb|HS| means a flag ``host-stack''.

\verb|is_mj| is used for selecting suitable \verb|CALL| definitions at
compilation time. This pseudo function returns \verb|true| under method-based
compilation context, otherwise \verb|false|. The host-stack styled definition should
be placed in the \verb|then| branch, and the user-stack styled definition is placed
in the \verb|else| branch as shown in the left of
Figure~\ref{code:stack_hybridization_meta_interp}. Then, the meta-hybrid \jit{}
compiler traces the \verb|then| branch in the context of
trace-based compilation, but traces the \verb|else| branch under method-based
compilation context.

\verb|US| and \verb|HS| mean trace- and method-based compilation contexts,
respectively. These special variables are used for detecting \jit{} compilation
context dynamically when executing \verb|RETURN| at runtime (not compilation time).

When defining \verb|CALL| in an interpreter, \verb|US| or \verb|HS| is
placed at the top of a user-stack when language developers define \verb|CALL|
instruction. At compilation time, these flags are treated as \emph{red}
variable, so an instruction pushing \verb|US| or \verb|HS| flag is left in a
resulting trace. The compiler also leaves the branching instruction (\texttt{if
  JIT\_flg == HS: ... else: ...}) in a resulting trace when tracing
\verb|RET|. This enables to find a \jit{} compilation context, and cooperate
resulting traces made from the different two strategies at runtime.

For example, there are two traces, one (\textit{A}) is made from trace-based
compilation and the other (\textit{B}) is from method-based compilation. When
a function call from \textit{A} to \textit{B} is occurred, a flag \verb|US|
is pushed to a user-defined stack. When executing a \verb|RET| instruction
in \textit{B}, the control executes a suitable definition by writing as shown in
the right of Figure~\ref{code:stack_hybridization_meta_interp}.

\begin{figure}[!t]
  \centering
  \begin{subfigure}[t]{0.49\columnwidth}
    \begin{minted}{python}
if instr == CALL:
 addr = bytecode[pc]
 # branch considering by
 # a JIT ctx.
 if is_mj():
  # push JIT flag (HS) to
  # ``user-stack''
  user_stack.push(HS)
  ret_val = interp(addr)
  user_stack.push(ret_value)
 else:
  # push JIT flag (US) to
  # ``user-stack''
  user_stack.push(US)
  user_stack.push(pc+1)
  pc = addr
     \end{minted}
   \end{subfigure}
  \begin{subfigure}[t]{0.49\columnwidth}
    \begin{minted}[firstnumber=last,]{python}
elif instr == RETURN:
 ret_val = user_stack.pop()
 # get JIT ctx. flag from
 # ``user-stack''
 JIT_flg = user_stack.pop()
 # check the JIT ctx. and branch
 if JIT_flg == HS:
  return ret_val
 else:
  ret_addr = user_stack.pop()
  user_stack.push(ret_val)
  pc = ret_addr
  \end{minted}
  \end{subfigure}
  \caption{A sketch of a interpreter definition with Stack Hybridization. Some hint
    functions (e.g., \texttt{can\_enter\_jit} and \texttt{jit\_merge\_point}), and other
    definitions are omitted for simplicity. \texttt{US} and \texttt{HS} represents
    user-stack and host-stack, respectively.}
  \label{code:stack_hybridization_meta_interp}
\end{figure}

\section{Evaluation}
\label{sec:evaluation}

In this section, we evaluate the basic performance of \baccaml{}'s trace-based and
method-based compilers. We first briefly introduce the current status of \baccaml{},
and how we took the data of microbenchmark programs. Next, we show the results of
evaluatation for \baccaml{} by running microbenchmark programs.

\subsection{Setup}
\label{sec:eval_setup}


\paragraph{Implementation}
\label{sec:eval_impl}

We implemented the \baccaml{} meta-hybrid \jit{} compiler framework based on the
\mincaml{} compiler~\cite{Sumii2005}. \mincaml{} is a small ML compiler designed for
education-purpose. MinCaml can generate native code almost as fast as other notable
compilers such as GCC or OCamlOpt. We did not extend \rpython{} itself because the
implementation of \rpython{} is too huge to comprehend. As an initial step, we
created a subset of \rpython{} on a compiler with reasonable implementation
size~\footnote{\baccaml{} itself is written in OCaml, and its implementation can be
  accessed at GitHub~(\url{https://github.com/prg-titech/BacCaml})}.

We also created a small functional programming language, namely \mincamlmm{}, with
the \baccaml{} framework for taking microbenchmark. It is almost same to \mincaml{},
but limited to unit, boolean and integer variables~\footnote{\mincamlmm{} is also
  available at GitHub~({\url{https://github.com/prg-titech/MinCaml})}}.

\paragraph{Methodology}
\label{sec:eval_method}

We attempted to run all of \mincaml{}'s test programs~\footnote{\url{https://github.com/esumii/min-caml/tree/master/test}} and
shootout~\footnote{\url{https://dada.perl.it/shootout/}} benchmark suite by
\mincamlmm{} and \baccaml{} before taking microbenchmark. Then, we selected all
programs that can be successfully worked in them. The names of microbenchmark
programs are shown in the X-axis of Figure~\ref{fig:bench}.

When taking microbenchmark, we set a threshold for starting \jit{} compilation at a
lower-than-normal value to simplify the situation. Basically, we set 100 as a
threshold to determine whether starting \jit{} compilation or not. Therefore, this
microbenchmark can arrive at a steady-state by attempting at most 50
iterations. Thus, we ran each program 150 times, and the first 50 trials were ignored
to exclude the warm-up.

Since \baccaml{} is a prototype, we convert a resulting trace to Assembly, compile it
by GCC at the compilation phase. The compiler then dispatches the control to the
machine code by using dynamic loading in the execution phase. Particularly,
trace-generation and compilation processes consume much time (approximately 80 \% of
a warm-up phase), so using a \jit{} native code generation framework such as
libgccjit~\footnote{\url{https://gcc.gnu.org/onlinedocs/jit/}} or GNU
Lightning~\footnote{\url{https://www.gnu.org/software/lightning/}} is left as
future work.

We ran all the microbenchmarks on Manjaro Linux with Linux kernel version
5.6.16-1-MANJARO and dedicated hardware with the following modules; CPU: AMD Ryzen 5
3600 6-Core Processor; Memory: 32 GB DDR4 2666Mhz PC4-21300.

In the Figure~\ref{fig:bench_jit_quality} and Figure~\ref{fig:bench_speedup},
\tikz{\draw[pattern=dots] rectangle (1ex,1.5ex);} means \baccaml{}'s tracing \jit{},
\tikz{\draw[fill=gray] rectangle (1ex,1.5ex)} means \baccaml{}'s method \jit{},
\tikz{\draw[pattern=crosshatch] rectangle (1ex,1.5ex)} means \baccaml{}'s
interpreter-only execution, \tikz{\draw[fill=white] rectangle(1ex,1.5ex);} means
MinCaml (AOT), and \tikz{\draw[pattern=sixpointed stars] rectangle(1ex,1.5ex);} means
\baccaml{}'s hybrid \jit{} (mixing tracing and method \jit{}).

\paragraph{Threats to Validity}
\label{par:threads_to_validity}

There are the following threats to validity in our evaluation (including the
experiment in the next section). Our method-based compilation is naive, then
the inference which trace-based compilation is faster than method-based compilation
has a possibility to be overturned by a full-fledged method-based compilers. This is
actually true not only to our method-based compiler, but also for our trace-based
compiler when compared against the state-of-the-art trace-based compilers like PyPy.

\subsection{Results of Standalone JIT Microbenchmark}
\label{sec:eval_microbenchmarking}

\begin{figure*}[!t]
  \centering
  \begin{subfigure}[t]{1.0\linewidth}
    \centering
    \includegraphics[width=.9\columnwidth]{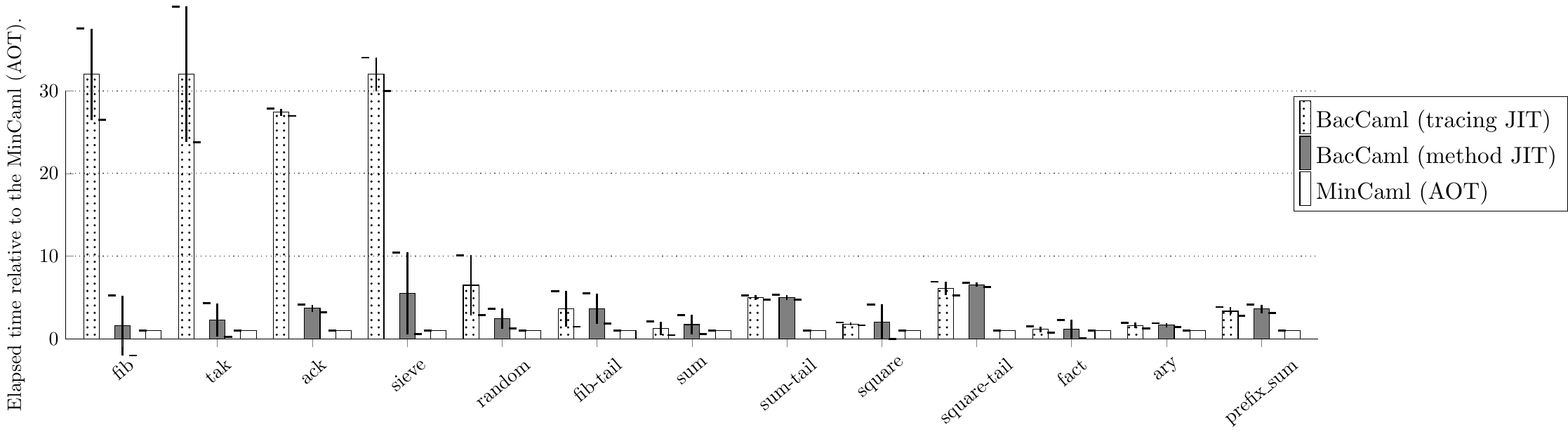}
    \caption{Elapsed time relative to the MinCaml (AOT) for each target
      program. Lower is better.}
    \label{fig:bench_jit_quality}
  \end{subfigure}
  \begin{subfigure}[t]{1.0\linewidth}
    \centering
    \includegraphics[width=.9\columnwidth]{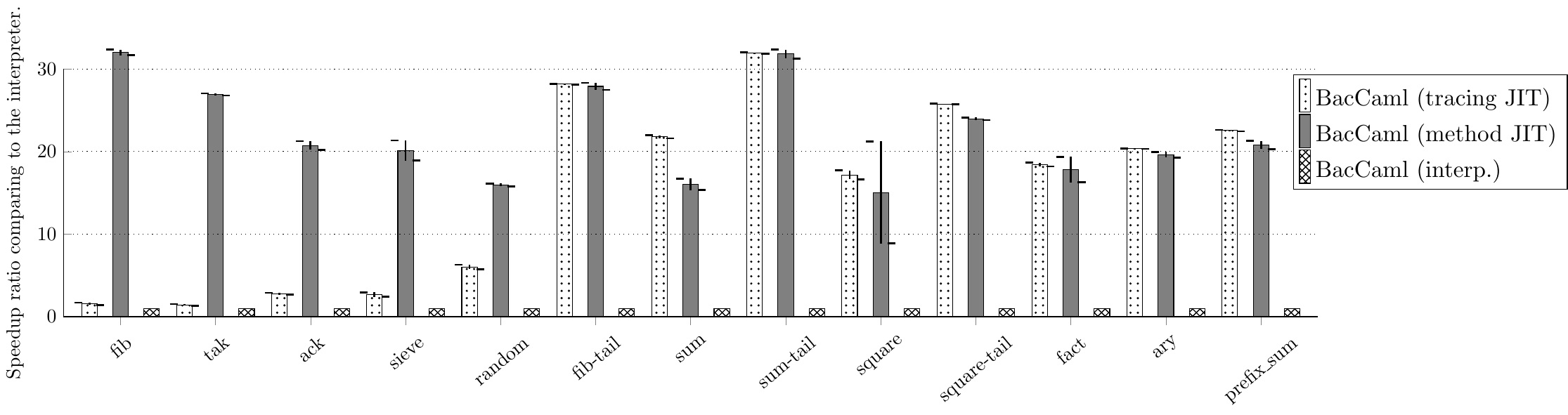}
    \caption{Speedup ratio of \jit{}s comparing to the interpreter-only
      execution for each target program. Higher is better.}
    \label{fig:bench_speedup}
  \end{subfigure}
  \caption{Results of standalone \jit{} microbenchmarking. The five programs on the
    left have a complex control flow, and the remaining programs have a straight
    control flow. The error bars represent the standard deviations.}
  \label{fig:bench}
\end{figure*}

For comparing the standalone performance of trace- and method-based compilation
strategies, we first applied both strategies separately for programs written in
\mincamlmm{}, and compared the performances of \mincamlmm{} with JIT with an
interpreter-only execution of \mincamlmm{} and the \mincaml{} ahead-of-time
compiler.

Before showing data, we explain the limitations of our method-based
compilation. Compared to other state-of-the-art method \jit{} compilers, our
method-based compilation does not inline them. It is because our method-based
compilation is designed to be applied to only programs with the path-divergence
problem. In other words, it is a fallback strategy when trace-based compilation does
not work well.

The results are shown in Figure~\ref{fig:bench_jit_quality}
and~\ref{fig:bench_speedup}. Note that Figure~\ref{fig:bench_jit_quality} is
normalized to the MinCaml (lower is better), but Figure~\ref{fig:bench_speedup} is to
the interpreter-only execution (higher is better).

Figure~\ref{fig:bench_jit_quality} illustrates the performances of the two \jit{}
compilations comparing to the elapsed time of \mincamlmm{} (AOT). In these results,
our trace-based compilation (\tikz{\draw[pattern=dots] rectangle (1ex,1.5ex);}) was
from 1.12 to 12.4x slower than \mincaml{} (AOT) (\tikz{\draw rectangle (1ex,1.5ex);})
in programs which have straight-forwarded control flow (fib-tail, sum, sum-tail,
square, square-tail, fact, ary, prefix\_sum). Our trace-based compilation was
effective on such programs since almost all executions are run on compiled
straight-line traces. However, it performs from 38.5 to 42.1x slower than other
strategies in programs with complex control flow (fib, ack, tak, sieve), since these
programs cause the path-divergence problem. In Figure~\ref{fig:bench_speedup}, our
trace-based compilation was from 6.02 to 31.92x faster than interpreter-only in
programs with straight-line control flow. However, it was still from 1.44 to 2.68x
faster than interpreter-only in programs with the path-divergence problem, since most
of the execution was done on the interpreter.

On the other hand, our method-based compilation (\tikz{\draw[fill=gray] rectangle
  (1ex,1.5ex)}) was from 1.16 to 6.52x slower than MinCaml (AOT) in
Figure~\ref{fig:bench_jit_quality}. Besides, from Figure~\ref{fig:bench_speedup}, our
method-based compilation also performs from 15.04 to 37.8x faster than
interpreter-only. From these results, most execution ran on a resulting trace. Our
strategy prevented the path-divergence problem since our method-based compilation
covered the entire body of a method.

Overall, our trace-based compilation was about 1.10x faster in programs with
straight-line control flow but about 11.8x slower in programs with complex control
flow than our method-based compilation. According to those results, we can say that
trace-based compilation's performance depends on the control flow of a target program
(when fitted to trace-based strategy, its performance was better than method-based
strategy). Still, a method-based strategy works well on average. Therefore, we argue
that combining the two strategies is vital for further speedup on \jit{}
compilation.




\section{Hybrid JIT Experiment}
\label{sec:experiment}

In this section, we demonstrate the result of an experiment for a hybrid \jit{}
compilation strategy. This experiment aims to confirm if there are programs that are
faster with a hybrid strategy than standalone strategies. This experiment was also
performed by the same implementations used in Section~\ref{sec:evaluation}.

\subsection{Setup}
\label{sec:experiment_setup}

\paragraph{Methodology}
\label{sec:experiment_method}

We first synthesized two types of functions, one is suitable for trace-based
compilation, and the other is for method-based compilation according to the result
shown in Section~\ref{sec:evaluation}. Then, we applied hybrid \jit{} compilation for
them; trace-based compilation is applied for a program with straight-line control
flow, and method-based compilation is applied for a program with complex control
flow. Finally, we compared the performance with standalone \jit{} strategies (tracing
\jit{} only and method \jit{} only) and \baccaml{}'s interpreter-only execution.

According to the result shown in Section~\ref{sec:eval_microbenchmarking}, we chose
sum, fib, and tak from the microbenchmark programs. It is because sum is faster in
trace-based compilation than in method-based compilation, and fib and tak are faster
in method-based compilation than a trace-based compilation. Then we manually
synthesized those functions for preparing test programs, namely sum-fib, fib-sum,
sum-tak and tak-sum, that shown in Figure~\ref{fig:experiment_prog}.

For taking data, we used the same implementations and hardware employed in
Section~\ref{sec:evaluation}. We also took 150 iterations and ignored the first 50
trials to exclude warm-up.

Since the algorithm that decides to apply which compilation strategy to which part of
a program is left for future work, we manually decided the program parts'
strategies. In a hybrid \jit{} compilation strategy, we applied trace-based
compilation to sum, and method-based compilation to fib and tak manually. Despite
this, in other strategies, we used only a single strategy for those test
programs. Specifically, we applied trace-based compilation to sum, fib and tak in a
tracing \jit{} only strategy, and applied method-based compilation for them in a
method \jit{} only strategy.

\begin{figure}[t]
  \centering
  \begin{subfigure}[t]{.48\linewidth}
    \begin{minted}[fontsize=\scriptsize]{ocaml}
let rec fib n =
  if n <= 1 then 1 else
    fib (n-1) + fib (n-2) in

let rec sum i n =
  if i <= 1 then n else
  let m = fib 10 in
    sum (i-1) (n+m) in

print_int (sum 1000 0)
    \end{minted}
    \caption{sum-fib}
    \end{subfigure}
  \begin{subfigure}[t]{.48\linewidth}
    \begin{minted}[fontsize=\scriptsize]{ocaml}
let rec sum acc n =
  if n <= 1 then acc else
    sum (acc+n) (n-1) in

let rec fib n =
  if n <= 2  then sum 0 1000
  else
    fib (n-1) + fib (n-2) in

print_int (fib 20)
     \end{minted}
     \caption{fib-sum}
   \end{subfigure}
   \begin{subfigure}[t]{.48\linewidth}
     \begin{minted}[fontsize=\scriptsize]{ocaml}
let rec tak x y z =
  if x <= y then z else
    tak (tak (x-1) y z)
        (tak (y-1) z x)
        (tak (z-1) x y) in

let rec sum i n =
  if i <= 1 then n else
  let m = tak 12 6 4 in
    sum (i-1) (n + m) in

print_int (sum 100 0)
     \end{minted}
     \caption{sum-tak}
   \end{subfigure}
   \begin{subfigure}[t]{.48\linewidth}
     \begin{minted}[fontsize=\scriptsize]{ocaml}
let rec sum i n =
  if i <= 1 then n else
    sum (i-1) (n+i) in

let rec tak x y z =
  if x <= y then sum 1000 0
  else
    tak (tak (x-1) y z)
        (tak (y-1) z x)
        (tak (z-1) x y) in

print_int (tak 8 4 2)
     \end{minted}
     \caption{tak-sum}
   \end{subfigure}
  \caption{Target programs written in \protect\mincamlmm{} used for the hybrid
    \protect\jit{} experiment.}
  \label{fig:experiment_prog}
\end{figure}

\subsection{Results of Hybrid JIT Experiment}
\label{sec:hybrid_jit_experiment}

\begin{figure}[!t]
  \centering
  \includegraphics[width=.85\columnwidth]{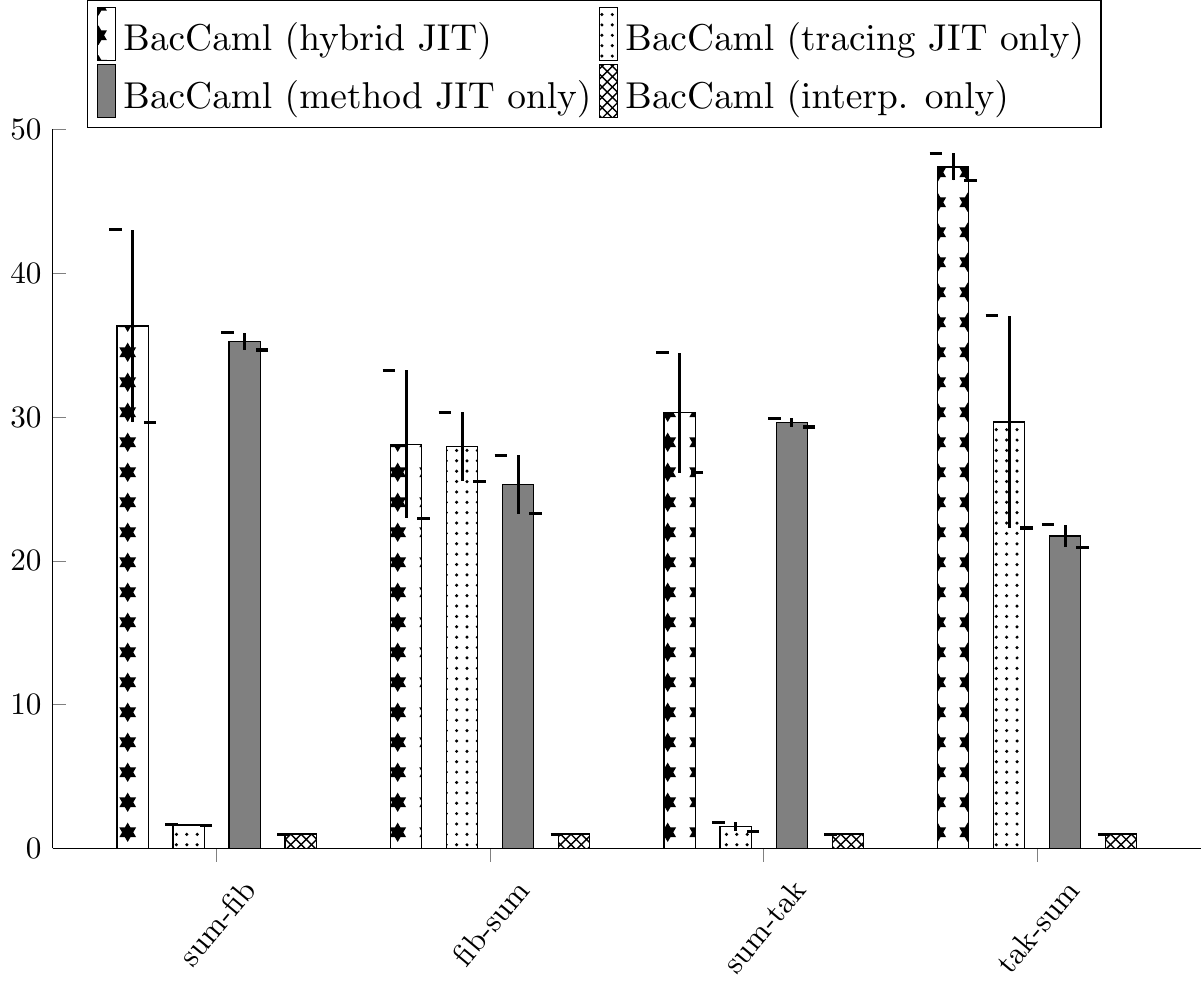}
  \caption{Results of hybrid \protect\jit{} microbenchmarking. X-axis represents the
    name of a target program, and Y-axis represents speedup ratio relative to the
    interpreter-only execution. Higher is better. The error bars represent the
    standard deviations.}
  \label{fig:bench_hybrid}
\end{figure}

The results of the hybrid \jit{} experiment are shown in
Figure~\ref{fig:bench_hybrid}. Overall, our hybrid compilation strategy
(\tikz{\draw[pattern=sixpointed stars] rectangle(1ex,1.5ex);}) was from 1.01 to 2.17x
faster than our method-based compilation-only. Our hybrid JIT strategy avoided the
overhead of recursive function calls in sum when executing the native code generated
from sum, since the recursive call part was inlined by trace-based compilation.

In contrast, our hybrid strategy was from 1.01 to 1.59x faster than the trace-based
compilation-only in fib-sum and tak-sum. Moreover, our hybrid strategy was about 20x
faster in sum-fib and sum-tak. This difference was caused by the structure of the
target program's control flow. In fib-sum and tak-sum, fib and tak can be connected
to the sum's recursive call and return parts. Otherwise, in sum-fib and sum-tak, sum
cannot be connected to entire fib and tak, since our trace-based compiler cannot
cover entire body of fib and tak functions by the path-divergence problem.

From the results, we can report that there are programs that can be run faster by the
hybrid compilation strategy.


\section{Related Work}
\label{sec:related_work}

\paragraph{Self-optimizing Abstract-syntax-tree interpreter.}

\sloppy
Self-optimizing abstract-syntax-tree interpreter~\cite{Wurthinger2012} also
enables language developers to implement effective virtual machines. The framework
and the compiler are called Truffle and Graal, respectively. The difference from our
system is the basic compilation unit. Our system is based on a meta-tracing compiler,
so the compilation unit is a trace. In contrust, Truffle/Graal applies partial
evaluation for an \astree{}-based interpreter of an interpreter at execution time. By
profiling the runtime types and values, it can optimize a base-program and run it
efficiently.


\paragraph{GraalSqueak.}

GraalSqueak~\cite{Niephaus:2019:GTS:3357390.3361024} is a Squeak/Smalltalk VM
implementation written in Truffle
framework. In~\cite{Niephaus:2018:GFS:3242947.3242948}, Niephaus et al. provided an
efficient way to compile a bytecode-formatted program; that is, they showed a way to
apply trace-based compilation with an AST-rewriting interpreter strategy.

We extend the meta-tracing \jit{} compilation framework to support method-based
compilation, but their approach involves creating an interpreter to enable
trace-based compilation on a partial evaluation-based meta-\jit{} compiler
framework. Their idea is to implement an interpreter with some specific hint
annotations to expand the loop of an application program. Sulong has already
demonstrated the same idea~\cite{Rigger:2016:BLL:2998415.2998416}, and it was applied
for implementing Squeak/Smalltalk \vm{}.


\paragraph{Region-based JIT compiler.}

HHVM~\cite{10.1145/3192366.3192374} is a high-performance VM for PHP and Hack
programming languages. An important aspect of HHVM 2nd generation is its region-based
\jit{} compiler. A region-based compiler~\cite{10.5555/225160.225189} is not
restricted to compile the entire body of methods, basic blocks, or straight-line
traces; it can compile a combination of several program areas. Their compilation
strategy is more flexible than our hybrid compilation strategy, because an HHVM
region-based compiler can compile basic blocks, the entire body of methods, loops,
and any combination of them. However, their approach is limited to a specific
language system; we aim to provide some flexibility of compilation as a meta \jit{}
compiler framework.

\paragraph{Lazy basic block versioning.}

Lazy basic block versioning~\cite{ChevalierBoisvert_et_al:LIPIcs:2015:5219} is a
\jit{} compilation technique based on basic blocks. This strategy combines type
specialization and code duplication to remove redundant type checking, and it
can generate effective machine code. Moreover, as well as HHVM's region-based
\jit{} compiler, it can compile straight-line code paths and the entire
method bodies by constructing basic blocks. Chevalier-Boisvert and Feely
implemented lazy basic block versioning on Higgs, a research-oriented
JavaScript \vm{}~\footnote{\url{https://github.com/higgsjs/Higgs}}.
Their method-based compilation is similar to ours; tracing both sides of
conditional branches, not inlining functions, and not handle loops specially.
The difference is that our method-based compilation is not based on basic
blocks, but on traces. Moreover, our hybrid compilation can be applied not only for
specific, but also for any languages.





\paragraph{HPS: High Performance Smalltalk.}

High Performance Smalltalk (HPS) is a virtual machine used in VisualWorks
Smalltalk~\cite{ViWorks}. HPS achieves efficient performance by well-planned stack
frame management. In HPS, the key technique for efficient implementation of contexts
is mapping (closure or method) activations to stack frames in runtime. HPS has three
context representaions. (1) \emph{Volatile} contexts: precedure activations which have
yet to be accessed as context objects. (2) \emph{Stable} contexts: the normal object
form of procedures. (3) \emph{Hybrid} contexts: a pair of a context object and its
associated activation. By preparing extra slots for hybrid contexts in the stack
frames, HPS can distinguish between hybrid and volatile contexts.
This technique is similar to our stack hybridization. Stack hybridization also has
the two contexts, which represent tracing and method \jit{}, respectively. Moreover,
the stack hybridization checks the return pc by a flag in an user-defined array
structure as well as HPS manages the context in a separate array object.
To avoid impacting the garbage collector, the header of a hybrid context is
spotted as an object including raw bits rather than object pointers.
However, stack hybridization is so naive that it currently does not consider the
impact of garbage collector.

\section{Conclusion and Future Work}
\label{sec:conclusion_futurework}

\subsection{Conclusion}
\label{sec:conclusion}

We proposed a meta-hybrid \jit{} compiler framework to take advantage of trace- and
method-based compilation strategies as a multilingual approach. For supporting the
idea, we chose a meta-tracing \jit{} compiler as our foundation and extended it to
perform method-based compilation using tracing. By customizing the following
features, we realized it: (1) trace entry/exit points, (2) conditional
branches, (3) function calls, and (4) loops. We also proposed Stack Hybridization: an
interpreter design to enable connecting native code generated from different
strategies. The key concept of Stack Hybridization is (1) embedding two types of
interpreter implementation styles into a single definition, (2) selecting an
appropriate style at just-in-time compilation time, and (3) putting a flag on the
stack data structure to indicate whether it is under trace- or method-based
compilation.

We implemented a prototype of our meta-hybrid \jit{} compiler framework called
\baccaml{} as a proof-of-concept. We created a small meta-tracing \jit{}
compiler on the \mincaml{} compiler, and supported method-based compilation by
extending trace-based compilation, and achieved Stack Hybridization on it.

We evaluated the basic performance of \baccaml{}'s trace- and method-based
compilers. The results showed that our trace-based compiler ran from 6.02 to 31.92x
faster than interpreter-only execution in programs with straight-line control flow,
but 1.44 to 2.68x faster in programs with complex control flow. Our method-based
compiler ran 15.04 to 37.8x faster than interpreter-only execution in all types of
programs.

We finally executed a synthetic experiment to confirm the usefulness of the hybrid
strategy, and reported that there are example programs that are faster with the
hybrid strategy.




\subsection{Future Work}
\label{sec:future_work}

\paragraph{Selecting a suitable strategy dynamically}

Since we focused on studying how to construct an essential mechanism of hybrid
compilation and how to connect code fragments generated from different strategies, we
have not investigated an approach for automatic selection of a suitable
strategy. Such a mechanism is needed for applying our idea to more complex and
productive programs. To realize it, we currently plan to create and combine the
following profiling and analyzing techniques: (1) Profiling runtime information
related to branching biases and the depth of function calls. If a target function has
branching bias or deeply-function calls, we apply trace-based compilation to it. On
the other hand, a target program has complex control flow; we apply method-based
compilation to it. (2) Statically analysis the structure of a target program. The
analyzer parses the program and examines the complexity of the target's control
flow. When it has straight-line control flow, we apply trace-based compilation for
it. On the other hand, if it has complex control flow, we apply a method-based
compilation on it.

\paragraph{Designing more fluent interpreter definition}

Making hybrid-capable interpreter definition easier is also our future work. To
support \sh{}, the language designer has to manually insert code fragments to
record/check different stack styles, which would be tedious and error-prone. This
would be resolved by providing annotation functions for function call/return. Those
annotation functions would also help develop further optimization techniques around
dynamic checks for hybridized stacks using stub return addresses.

\paragraph{Realizing our idea at a production level}

\baccaml{} is just a proof-of-concept meta-JIT compiler framework; therefore, to
measure the impact of our hybrid \jit{} compilation on real-world applications, we
will create our hybrid compilation strategy on a practical framework, such as
\rpython{} or Truffle/Graal.

\begin{acks}
  We would like to thank Stefan Marr, Carl Friedrich Bolz, and Fabio Niephaus for
  their comments on earlier versions of the paper.  We also would like to thank the
  members of the Programming Research Group at Tokyo Institute of Technology for
  their comments and suggestions.
  This work was supported by KAKENHI (18H03219).
\end{acks}



\begin{thebibliography}{38}


\ifx \showCODEN    \undefined \def \showCODEN     #1{\unskip}     \fi
\ifx \showDOI      \undefined \def \showDOI       #1{#1}\fi
\ifx \showISBNx    \undefined \def \showISBNx     #1{\unskip}     \fi
\ifx \showISBNxiii \undefined \def \showISBNxiii  #1{\unskip}     \fi
\ifx \showISSN     \undefined \def \showISSN      #1{\unskip}     \fi
\ifx \showLCCN     \undefined \def \showLCCN      #1{\unskip}     \fi
\ifx \shownote     \undefined \def \shownote      #1{#1}          \fi
\ifx \showarticletitle \undefined \def \showarticletitle #1{#1}   \fi
\ifx \showURL      \undefined \def \showURL       {\relax}        \fi
\providecommand\bibfield[2]{#2}
\providecommand\bibinfo[2]{#2}
\providecommand\natexlab[1]{#1}
\providecommand\showeprint[2][]{arXiv:#2}

\bibitem[\protect\citeauthoryear{Adams, Evans, Maher, Ottoni, Paroski, Simmers,
  Smith, and Facebook}{Adams et~al\mbox{.}}{2014}]%
        {Adams2014}
\bibfield{author}{\bibinfo{person}{Keith Adams}, \bibinfo{person}{Jason Evans},
  \bibinfo{person}{Bertrand Maher}, \bibinfo{person}{Guilherme Ottoni},
  \bibinfo{person}{Andrew Paroski}, \bibinfo{person}{Brett Simmers},
  \bibinfo{person}{Edwin Smith}, {and} \bibinfo{person}{Owen~Yamauchi
  Facebook}.} \bibinfo{year}{2014}\natexlab{}.
\newblock \showarticletitle{The HipHop Virtual Machine}. In
  \bibinfo{booktitle}{\emph{Proceedings of the 2014 ACM International
  Conference on Object Oriented Programming Systems Languages \& Applications}}
  (Portland, Oregon, USA) \emph{(\bibinfo{series}{OOPSLA '14})}.
  \bibinfo{pages}{777--790}.
\newblock
\showISBNx{9781450325851}
\urldef\tempurl%
\url{https://doi.org/10.1145/2660193.2660199}
\showDOI{\tempurl}


\bibitem[\protect\citeauthoryear{Arnold, Fink, Grove, Hind, and Sweeney}{Arnold
  et~al\mbox{.}}{2000}]%
        {IBM:10.1145/353171.353175}
\bibfield{author}{\bibinfo{person}{Matthew Arnold}, \bibinfo{person}{Stephen
  Fink}, \bibinfo{person}{David Grove}, \bibinfo{person}{Michael Hind}, {and}
  \bibinfo{person}{Peter~F. Sweeney}.} \bibinfo{year}{2000}\natexlab{}.
\newblock \showarticletitle{Adaptive Optimization in the Jalape\~{n}o JVM}. In
  \bibinfo{booktitle}{\emph{Proceedings of the 15th ACM SIGPLAN Conference on
  Object-Oriented Programming, Systems, Languages, and Applications}}
  (Minneapolis, Minnesota, USA) \emph{(\bibinfo{series}{OOPSLA '00})}.
  \bibinfo{publisher}{Association for Computing Machinery},
  \bibinfo{address}{New York, NY, USA}, \bibinfo{pages}{47–65}.
\newblock
\showISBNx{158113200X}
\urldef\tempurl%
\url{https://doi.org/10.1145/353171.353175}
\showDOI{\tempurl}


\bibitem[\protect\citeauthoryear{Bala, Duesterwald, and Banerjia}{Bala
  et~al\mbox{.}}{2000}]%
        {Bala2000}
\bibfield{author}{\bibinfo{person}{Vasanth Bala}, \bibinfo{person}{Evelyn
  Duesterwald}, {and} \bibinfo{person}{Sanjeev Banerjia}.}
  \bibinfo{year}{2000}\natexlab{}.
\newblock \showarticletitle{{Dynamo: a Transparent Dynamic Optimization
  System}}. In \bibinfo{booktitle}{\emph{Proceedings of the ACM SIGPLAN 2000
  Conference on Programming Language Design and Implementation}}.
\newblock
\showISBNx{1-58113-199-2}
\showISSN{03621340}
\urldef\tempurl%
\url{https://doi.org/10.1145/349299.349303}
\showDOI{\tempurl}
\showeprint{1003.4074}


\bibitem[\protect\citeauthoryear{Bauman, Bolz, Hirschfeld, Kirilichev, Pape,
  Siek, and Tobin-Hochstadt}{Bauman et~al\mbox{.}}{2015a}]%
        {Bauman:2015:PTJ:2784731.2784740}
\bibfield{author}{\bibinfo{person}{Spenser Bauman},
  \bibinfo{person}{Carl~Friedrich Bolz}, \bibinfo{person}{Robert Hirschfeld},
  \bibinfo{person}{Vasily Kirilichev}, \bibinfo{person}{Tobias Pape},
  \bibinfo{person}{Jeremy~G. Siek}, {and} \bibinfo{person}{Sam
  Tobin-Hochstadt}.} \bibinfo{year}{2015}\natexlab{a}.
\newblock \showarticletitle{Pycket: A Tracing JIT for a Functional Language}.
  In \bibinfo{booktitle}{\emph{Proceedings of the 20th ACM SIGPLAN
  International Conference on Functional Programming}} (Vancouver, BC, Canada)
  \emph{(\bibinfo{series}{ICFP 2015})}. \bibinfo{publisher}{ACM},
  \bibinfo{address}{New York, NY, USA}, \bibinfo{pages}{22--34}.
\newblock
\showISBNx{978-1-4503-3669-7}
\urldef\tempurl%
\url{https://doi.org/10.1145/2784731.2784740}
\showDOI{\tempurl}


\bibitem[\protect\citeauthoryear{Bauman, Bolz, Hirschfeld, Kirilichev, Pape,
  Siek, and Tobin-Hochstadt}{Bauman et~al\mbox{.}}{2015b}]%
        {PycketInterp}
\bibfield{author}{\bibinfo{person}{Spenser Bauman},
  \bibinfo{person}{Carl~Friedrich Bolz}, \bibinfo{person}{Robert Hirschfeld},
  \bibinfo{person}{Vasily Kirilichev}, \bibinfo{person}{Tobias Pape},
  \bibinfo{person}{Jeremy~G. Siek}, {and} \bibinfo{person}{Sam
  Tobin-Hochstadt}.} \bibinfo{year}{2015}\natexlab{b}.
\newblock \bibinfo{title}{Pycket's Interpreter Definition}.
\newblock
\newblock
\newblock
\shownote{\url{https://github.com/pycket/pycket/blob/master/pycket/interpreter.py\#L2505},
  visited 2020-09-07.}


\bibitem[\protect\citeauthoryear{Bebenita, Brandner, Fahndrich, Logozzo,
  Schulte, Tillmann, and Venter}{Bebenita et~al\mbox{.}}{2010}]%
        {Bebenita2010}
\bibfield{author}{\bibinfo{person}{Michael Bebenita}, \bibinfo{person}{Florian
  Brandner}, \bibinfo{person}{Manuel Fahndrich}, \bibinfo{person}{Francesco
  Logozzo}, \bibinfo{person}{Wolfram Schulte}, \bibinfo{person}{Nikolai
  Tillmann}, {and} \bibinfo{person}{Herman Venter}.}
  \bibinfo{year}{2010}\natexlab{}.
\newblock \showarticletitle{SPUR: A Trace-based JIT Compiler for CIL}. In
  \bibinfo{booktitle}{\emph{Proceedings of the ACM International Conference on
  Object Oriented Programming Systems Languages and Applications}} (Reno/Tahoe,
  Nevada, USA) \emph{(\bibinfo{series}{OOPSLA '10})}. \bibinfo{publisher}{ACM},
  \bibinfo{address}{New York, NY, USA}, \bibinfo{pages}{708--725}.
\newblock
\showISBNx{978-1-4503-0203-6}
\urldef\tempurl%
\url{https://doi.org/10.1145/1869459.1869517}
\showDOI{\tempurl}


\bibitem[\protect\citeauthoryear{Bolz, Cuni, FijaBkowski, Leuschel, Pedroni,
  and Rigo}{Bolz et~al\mbox{.}}{2011a}]%
        {Bolz:2011:ARP:1929501.1929508}
\bibfield{author}{\bibinfo{person}{Carl~Friedrich Bolz},
  \bibinfo{person}{Antonio Cuni}, \bibinfo{person}{Maciej FijaBkowski},
  \bibinfo{person}{Michael Leuschel}, \bibinfo{person}{Samuele Pedroni}, {and}
  \bibinfo{person}{Armin Rigo}.} \bibinfo{year}{2011}\natexlab{a}.
\newblock \showarticletitle{Allocation Removal by Partial Evaluation in a
  Tracing JIT}. In \bibinfo{booktitle}{\emph{Proceedings of the 20th ACM
  SIGPLAN Workshop on Partial Evaluation and Program Manipulation}} (Austin,
  Texas, USA) \emph{(\bibinfo{series}{PEPM '11})}. \bibinfo{publisher}{ACM},
  \bibinfo{address}{New York, NY, USA}, \bibinfo{pages}{43--52}.
\newblock
\showISBNx{978-1-4503-0485-6}
\urldef\tempurl%
\url{https://doi.org/10.1145/1929501.1929508}
\showDOI{\tempurl}


\bibitem[\protect\citeauthoryear{Bolz, Cuni, FijaBkowski, Leuschel, Pedroni,
  and Rigo}{Bolz et~al\mbox{.}}{2011b}]%
        {Bolz:2011:RFM:2069172.2069181}
\bibfield{author}{\bibinfo{person}{Carl~Friedrich Bolz},
  \bibinfo{person}{Antonio Cuni}, \bibinfo{person}{Maciej FijaBkowski},
  \bibinfo{person}{Michael Leuschel}, \bibinfo{person}{Samuele Pedroni}, {and}
  \bibinfo{person}{Armin Rigo}.} \bibinfo{year}{2011}\natexlab{b}.
\newblock \showarticletitle{Runtime Feedback in a Meta-tracing JIT for
  Efficient Dynamic Languages}. In \bibinfo{booktitle}{\emph{Proceedings of the
  6th Workshop on Implementation, Compilation, Optimization of Object-Oriented
  Languages, Programs and Systems}} (Lancaster, United Kingdom)
  \emph{(\bibinfo{series}{ICOOOLPS '11})}. \bibinfo{publisher}{ACM},
  \bibinfo{address}{New York, NY, USA}, Article \bibinfo{articleno}{9},
  \bibinfo{numpages}{8}~pages.
\newblock
\showISBNx{978-1-4503-0894-6}
\urldef\tempurl%
\url{https://doi.org/10.1145/2069172.2069181}
\showDOI{\tempurl}


\bibitem[\protect\citeauthoryear{Bolz, Cuni, Fijalkowski, and Rigo}{Bolz
  et~al\mbox{.}}{2009}]%
        {Bolz2009}
\bibfield{author}{\bibinfo{person}{Carl~Friedrich Bolz},
  \bibinfo{person}{Antonio Cuni}, \bibinfo{person}{Maciej Fijalkowski}, {and}
  \bibinfo{person}{Armin Rigo}.} \bibinfo{year}{2009}\natexlab{}.
\newblock \showarticletitle{Tracing the Meta-level: PyPy's Tracing JIT
  Compiler}. In \bibinfo{booktitle}{\emph{Proceedings of the 4th Workshop on
  the Implementation, Compilation, Optimization of Object-Oriented Languages
  and Programming Systems}} (Genova, Italy). \bibinfo{publisher}{ACM},
  \bibinfo{address}{New York, NY, USA}, \bibinfo{pages}{18--25}.
\newblock
\showISBNx{978-1-60558-541-3}
\urldef\tempurl%
\url{https://doi.org/10.1145/1565824.1565827}
\showDOI{\tempurl}


\bibitem[\protect\citeauthoryear{Bolz and Tratt}{Bolz and Tratt}{2015}]%
        {BOLZ2015408}
\bibfield{author}{\bibinfo{person}{Carl~Friedrich Bolz} {and}
  \bibinfo{person}{Laurence Tratt}.} \bibinfo{year}{2015}\natexlab{}.
\newblock \showarticletitle{{{The Impact of Meta-tracing on VM Design and
  Implementation}}}.
\newblock \bibinfo{journal}{\emph{Science of Computer Programming}}
  \bibinfo{volume}{98} (\bibinfo{year}{2015}), \bibinfo{pages}{408 -- 421}.
\newblock
\showISSN{0167-6423}
\urldef\tempurl%
\url{https://doi.org/10.1016/j.scico.2013.02.001}
\showDOI{\tempurl}
\newblock
\shownote{Special Issue on Advances in Dynamic Languages.}


\bibitem[\protect\citeauthoryear{Chevalier-Boisvert and
  Feeley}{Chevalier-Boisvert and Feeley}{2015}]%
        {ChevalierBoisvert_et_al:LIPIcs:2015:5219}
\bibfield{author}{\bibinfo{person}{Maxime Chevalier-Boisvert} {and}
  \bibinfo{person}{Marc Feeley}.} \bibinfo{year}{2015}\natexlab{}.
\newblock \showarticletitle{{Simple and Effective Type Check Removal through
  Lazy Basic Block Versioning}}. In \bibinfo{booktitle}{\emph{29th European
  Conference on Object-Oriented Programming (ECOOP15)}}
  \emph{(\bibinfo{series}{Leibniz International Proceedings in Informatics
  (LIPIcs)}, Vol.~\bibinfo{volume}{37})},
  \bibfield{editor}{\bibinfo{person}{John~Tang Boyland}} (Ed.).
  \bibinfo{publisher}{Schloss Dagstuhl--Leibniz-Zentrum fuer Informatik},
  \bibinfo{address}{Dagstuhl, Germany}, \bibinfo{pages}{101--123}.
\newblock
\showISBNx{978-3-939897-86-6}
\showISSN{1868-8969}
\urldef\tempurl%
\url{https://doi.org/10.4230/LIPIcs.ECOOP.2015.101}
\showDOI{\tempurl}


\bibitem[\protect\citeauthoryear{Deutsch and Schiffman}{Deutsch and
  Schiffman}{1984}]%
        {Smalltalk80:10.1145/800017.800542}
\bibfield{author}{\bibinfo{person}{L.~Peter Deutsch} {and}
  \bibinfo{person}{Allan~M. Schiffman}.} \bibinfo{year}{1984}\natexlab{}.
\newblock \showarticletitle{Efficient Implementation of the Smalltalk-80
  System}. In \bibinfo{booktitle}{\emph{Proceedings of the 11th ACM
  SIGACT-SIGPLAN Symposium on Principles of Programming Languages}} (Salt Lake
  City, Utah, USA) \emph{(\bibinfo{series}{POPL '84})}.
  \bibinfo{publisher}{Association for Computing Machinery},
  \bibinfo{address}{New York, NY, USA}, \bibinfo{pages}{297--302}.
\newblock
\showISBNx{0897911253}
\urldef\tempurl%
\url{https://doi.org/10.1145/800017.800542}
\showDOI{\tempurl}


\bibitem[\protect\citeauthoryear{Felgentreff, Pape, Rein, and
  Hirschfeld}{Felgentreff et~al\mbox{.}}{2016}]%
        {10.1145/2991041.2991062}
\bibfield{author}{\bibinfo{person}{Tim Felgentreff}, \bibinfo{person}{Tobias
  Pape}, \bibinfo{person}{Patrick Rein}, {and} \bibinfo{person}{Robert
  Hirschfeld}.} \bibinfo{year}{2016}\natexlab{}.
\newblock \showarticletitle{How to Build a High-Performance VM for
  Squeak/Smalltalk in Your Spare Time: An Experience Report of Using the
  RPython Toolchain}. In \bibinfo{booktitle}{\emph{Proceedings of the 11th
  Edition of the International Workshop on Smalltalk Technologies}} (Prague,
  Czech Republic) \emph{(\bibinfo{series}{IWST '16})}.
  \bibinfo{publisher}{Association for Computing Machinery},
  \bibinfo{address}{New York, NY, USA}, Article \bibinfo{articleno}{21},
  \bibinfo{numpages}{10}~pages.
\newblock
\showISBNx{9781450345248}
\urldef\tempurl%
\url{https://doi.org/10.1145/2991041.2991062}
\showDOI{\tempurl}


\bibitem[\protect\citeauthoryear{Foundation}{Foundation}{2007}]%
        {llvm}
\bibfield{author}{\bibinfo{person}{LLVM Foundation}.}
  \bibinfo{year}{2007}\natexlab{}.
\newblock \bibinfo{title}{The LLVM Compiler Infrastucture}.
\newblock
\newblock
\urldef\tempurl%
\url{https://llvm.org/}
\showURL{%
\tempurl}


\bibitem[\protect\citeauthoryear{Gal, Eich, Shaver, Anderson, Mandelin,
  Haghighat, Kaplan, Hoare, Zbarsky, Orendorff, Ruderman, Smith, Reitmaier,
  Bebenita, Chang, and Franz}{Gal et~al\mbox{.}}{2009}]%
        {Gal2009}
\bibfield{author}{\bibinfo{person}{Andreas Gal}, \bibinfo{person}{Brendan
  Eich}, \bibinfo{person}{Mike Shaver}, \bibinfo{person}{David Anderson},
  \bibinfo{person}{David Mandelin}, \bibinfo{person}{Mohammad~R. Haghighat},
  \bibinfo{person}{Blake Kaplan}, \bibinfo{person}{Graydon Hoare},
  \bibinfo{person}{Boris Zbarsky}, \bibinfo{person}{Jason Orendorff},
  \bibinfo{person}{Jesse Ruderman}, \bibinfo{person}{Edwin~W. Smith},
  \bibinfo{person}{Rick Reitmaier}, \bibinfo{person}{Michael Bebenita},
  \bibinfo{person}{Mason Chang}, {and} \bibinfo{person}{Michael Franz}.}
  \bibinfo{year}{2009}\natexlab{}.
\newblock \showarticletitle{{Trace-Based Just-in-Time Type Specialization for
  Dynamic Languages}}. In \bibinfo{booktitle}{\emph{Proceedings of the 30th ACM
  SIGPLAN Conference on Programming Language Design and Implementation}}
  (Dublin, Ireland) \emph{(\bibinfo{series}{PLDI '09})}.
  \bibinfo{publisher}{Association for Computing Machinery},
  \bibinfo{address}{New York, NY, USA}, \bibinfo{pages}{465–478}.
\newblock
\showISBNx{9781605583921}
\urldef\tempurl%
\url{https://doi.org/10.1145/1542476.1542528}
\showDOI{\tempurl}


\bibitem[\protect\citeauthoryear{Gal, Probst, and Franz}{Gal
  et~al\mbox{.}}{2006}]%
        {Gal:2006:10.1145/1134760.1134780}
\bibfield{author}{\bibinfo{person}{Andreas Gal}, \bibinfo{person}{Christian~W.
  Probst}, {and} \bibinfo{person}{Michael Franz}.}
  \bibinfo{year}{2006}\natexlab{}.
\newblock \showarticletitle{HotpathVM: An Effective JIT Compiler for
  Resource-Constrained Devices}. In \bibinfo{booktitle}{\emph{Proceedings of
  the 2nd International Conference on Virtual Execution Environments}} (Ottawa,
  Ontario, Canada) \emph{(\bibinfo{series}{VEE '06})}.
  \bibinfo{publisher}{Association for Computing Machinery},
  \bibinfo{address}{New York, NY, USA}, \bibinfo{pages}{144–153}.
\newblock
\showISBNx{1595933328}
\urldef\tempurl%
\url{https://doi.org/10.1145/1134760.1134780}
\showDOI{\tempurl}


\bibitem[\protect\citeauthoryear{Google}{Google}{2015}]%
        {googlev8}
\bibfield{author}{\bibinfo{person}{Google}.} \bibinfo{year}{2015}\natexlab{}.
\newblock \bibinfo{title}{Google’s High-performance Open Source JavaScript
  and WebAssembly Engine.}
\newblock
\newblock
\newblock
\shownote{\url{https://v8.dev/}, visited 2020-10-16.}


\bibitem[\protect\citeauthoryear{Hank, Hwu, and Rau}{Hank
  et~al\mbox{.}}{1995}]%
        {10.5555/225160.225189}
\bibfield{author}{\bibinfo{person}{Richard~E. Hank},
  \bibinfo{person}{Wen-Mei~W. Hwu}, {and} \bibinfo{person}{B.~Ramakrishna
  Rau}.} \bibinfo{year}{1995}\natexlab{}.
\newblock \showarticletitle{Region-Based Compilation: An Introduction and
  Motivation}. In \bibinfo{booktitle}{\emph{Proceedings of the 28th Annual
  International Symposium on Microarchitecture}} (Ann Arbor, Michigan, USA)
  \emph{(\bibinfo{series}{MICRO 28})}. \bibinfo{publisher}{IEEE Computer
  Society Press}, \bibinfo{address}{Washington, DC, USA},
  \bibinfo{pages}{158--168}.
\newblock
\showISBNx{0818673494}


\bibitem[\protect\citeauthoryear{Haupt, Hirschfeld, Pape, Gabrysiak, Marr,
  Bergmann, Heise, Kleine, and Krahn}{Haupt et~al\mbox{.}}{2010}]%
        {Haupt:2010:SFV:1822090.1822098}
\bibfield{author}{\bibinfo{person}{Michael Haupt}, \bibinfo{person}{Robert
  Hirschfeld}, \bibinfo{person}{Tobias Pape}, \bibinfo{person}{Gregor
  Gabrysiak}, \bibinfo{person}{Stefan Marr}, \bibinfo{person}{Arne Bergmann},
  \bibinfo{person}{Arvid Heise}, \bibinfo{person}{Matthias Kleine}, {and}
  \bibinfo{person}{Robert Krahn}.} \bibinfo{year}{2010}\natexlab{}.
\newblock \showarticletitle{The SOM Family: Virtual Machines for Teaching and
  Research}. In \bibinfo{booktitle}{\emph{Proceedings of the Fifteenth Annual
  Conference on Innovation and Technology in Computer Science Education}}
  (Bilkent, Ankara, Turkey) \emph{(\bibinfo{series}{ITiCSE '10})}.
  \bibinfo{publisher}{ACM}, \bibinfo{address}{New York, NY, USA},
  \bibinfo{pages}{18--22}.
\newblock
\showISBNx{978-1-60558-820-9}
\urldef\tempurl%
\url{https://doi.org/10.1145/1822090.1822098}
\showDOI{\tempurl}


\bibitem[\protect\citeauthoryear{Hayashizaki, Wu, Inoue, Serrano, and
  Nakatani}{Hayashizaki et~al\mbox{.}}{2011}]%
        {Hayashizaki:2011:IPT:1950365.1950412}
\bibfield{author}{\bibinfo{person}{Hiroshige Hayashizaki},
  \bibinfo{person}{Peng Wu}, \bibinfo{person}{Hiroshi Inoue},
  \bibinfo{person}{Mauricio~J. Serrano}, {and} \bibinfo{person}{Toshio
  Nakatani}.} \bibinfo{year}{2011}\natexlab{}.
\newblock \showarticletitle{Improving the Performance of Trace-based Systems by
  False Loop Filtering}. In \bibinfo{booktitle}{\emph{Proceedings of the
  Sixteenth International Conference on Architectural Support for Programming
  Languages and Operating Systems}} (Newport Beach, California, USA)
  \emph{(\bibinfo{series}{ASPLOS XVI})}. \bibinfo{publisher}{ACM},
  \bibinfo{address}{New York, NY, USA}, \bibinfo{pages}{405--418}.
\newblock
\showISBNx{978-1-4503-0266-1}
\urldef\tempurl%
\url{https://doi.org/10.1145/1950365.1950412}
\showDOI{\tempurl}


\bibitem[\protect\citeauthoryear{Huang, Masuhara, and Aotani}{Huang
  et~al\mbox{.}}{2016a}]%
        {Huang2016}
\bibfield{author}{\bibinfo{person}{Ruochen Huang}, \bibinfo{person}{Hidehiko
  Masuhara}, {and} \bibinfo{person}{Tomoyuki Aotani}.}
  \bibinfo{year}{2016}\natexlab{a}.
\newblock \showarticletitle{Improving Sequential Performance of Erlang Based on
  a Meta-tracing Just-In-Time Compiler}. In
  \bibinfo{booktitle}{\emph{International Symposium on Trends in Functional
  Programming}}. Springer, \bibinfo{pages}{44--58}.
\newblock


\bibitem[\protect\citeauthoryear{Huang, Masuhara, and Aotani}{Huang
  et~al\mbox{.}}{2016b}]%
        {PyrlangInterp}
\bibfield{author}{\bibinfo{person}{Rouchen Huang}, \bibinfo{person}{Hidehiko
  Masuhara}, {and} \bibinfo{person}{Tomoyuki Aotani}.}
  \bibinfo{year}{2016}\natexlab{b}.
\newblock \bibinfo{title}{Pyrlang's Interpreter Definition}.
\newblock
\newblock
\newblock
\shownote{\url{https://bitbucket.org/hrc706/pyrlang/src/0d4fa6b4d7e6a78d8abece9cdbdc38806ef819cd/interpreter/interp.py\#lines-666},
  visited 2020-09-07.}


\bibitem[\protect\citeauthoryear{Inoue, Hayashizaki, Wu, and Nakatani}{Inoue
  et~al\mbox{.}}{2011}]%
        {Inoue2011}
\bibfield{author}{\bibinfo{person}{Hiroshi Inoue}, \bibinfo{person}{Hiroshige
  Hayashizaki}, \bibinfo{person}{Peng Wu}, {and} \bibinfo{person}{Toshio
  Nakatani}.} \bibinfo{year}{2011}\natexlab{}.
\newblock \showarticletitle{{A trace-based Java JIT Compiler Retrofitted from a
  Method-based Compiler}}.
\newblock \bibinfo{journal}{\emph{Proceedings - International Symposium on Code
  Generation and Optimization}}, \bibinfo{pages}{246--256}.
\newblock
\showISBNx{9781612843551}
\showISSN{03621340}
\urldef\tempurl%
\url{https://doi.org/10.1109/CGO.2011.5764692}
\showDOI{\tempurl}


\bibitem[\protect\citeauthoryear{Marr and Ducasse}{Marr and Ducasse}{2015}]%
        {Marr:2015:TVP:2814270.2814275}
\bibfield{author}{\bibinfo{person}{Stefan Marr} {and}
  \bibinfo{person}{St{\'e}phane Ducasse}.} \bibinfo{year}{2015}\natexlab{}.
\newblock \showarticletitle{Tracing vs. Partial Evaluation: Comparing
  Meta-compilation Approaches for Self-optimizing Interpreters}. In
  \bibinfo{booktitle}{\emph{Proceedings of the 2015 ACM SIGPLAN International
  Conference on Object-Oriented Programming, Systems, Languages, and
  Applications}} (Pittsburgh, PA, USA) \emph{(\bibinfo{series}{OOPSLA '15})}.
  \bibinfo{publisher}{ACM}, \bibinfo{address}{New York, NY, USA},
  \bibinfo{pages}{821--839}.
\newblock
\showISBNx{978-1-4503-3689-5}
\urldef\tempurl%
\url{https://doi.org/10.1145/2814270.2814275}
\showDOI{\tempurl}


\bibitem[\protect\citeauthoryear{Miranda}{Miranda}{1999}]%
        {ViWorks}
\bibfield{author}{\bibinfo{person}{Eliot Miranda}.}
  \bibinfo{year}{1999}\natexlab{}.
\newblock \showarticletitle{Context Management in VisualWorks 5i}. In
  \bibinfo{booktitle}{\emph{OOPSLA'99 Workshop on Simplicity, Performance and
  Portability in Virtual Machine Design}}. \bibinfo{address}{Denver, CO}.
\newblock
\urldef\tempurl%
\url{http://www.esug.org/data/Articles/misc/oopsla99-contexts.pdf}
\showURL{%
\tempurl}


\bibitem[\protect\citeauthoryear{Mozilla}{Mozilla}{2016}]%
        {ionmonkey}
\bibfield{author}{\bibinfo{person}{Mozilla}.} \bibinfo{year}{2016}\natexlab{}.
\newblock \bibinfo{title}{{IonMonkey, the Next Generation JavaScript JIT for
  SpiderMonkey}}.
\newblock
\newblock
\urldef\tempurl%
\url{https://wiki.mozilla.org/IonMonkey}
\showURL{%
\tempurl}


\bibitem[\protect\citeauthoryear{Niephaus, Felgentreff, and
  Hirschfeld}{Niephaus et~al\mbox{.}}{2018}]%
        {Niephaus:2018:GFS:3242947.3242948}
\bibfield{author}{\bibinfo{person}{Fabio Niephaus}, \bibinfo{person}{Tim
  Felgentreff}, {and} \bibinfo{person}{Robert Hirschfeld}.}
  \bibinfo{year}{2018}\natexlab{}.
\newblock \showarticletitle{GraalSqueak: A Fast Smalltalk Bytecode Interpreter
  Written in an AST Interpreter Framework}. In
  \bibinfo{booktitle}{\emph{Proceedings of the 13th Workshop on Implementation,
  Compilation, Optimization of Object-Oriented Languages, Programs and
  Systems}} (Amsterdam, Netherlands) \emph{(\bibinfo{series}{ICOOOLPS '18})}.
  \bibinfo{publisher}{ACM}, \bibinfo{address}{New York, NY, USA},
  \bibinfo{pages}{30--35}.
\newblock
\showISBNx{978-1-4503-5804-0}
\urldef\tempurl%
\url{https://doi.org/10.1145/3242947.3242948}
\showDOI{\tempurl}


\bibitem[\protect\citeauthoryear{Niephaus, Felgentreff, and
  Hirschfeld}{Niephaus et~al\mbox{.}}{2019}]%
        {Niephaus:2019:GTS:3357390.3361024}
\bibfield{author}{\bibinfo{person}{Fabio Niephaus}, \bibinfo{person}{Tim
  Felgentreff}, {and} \bibinfo{person}{Robert Hirschfeld}.}
  \bibinfo{year}{2019}\natexlab{}.
\newblock \showarticletitle{GraalSqueak: Toward a Smalltalk-based Tooling
  Platform for Polyglot Programming}. In \bibinfo{booktitle}{\emph{Proceedings
  of the 16th ACM SIGPLAN International Conference on Managed Programming
  Languages and Runtimes}} (Athens, Greece) \emph{(\bibinfo{series}{MPLR
  '19})}. \bibinfo{publisher}{ACM}, \bibinfo{address}{New York, NY, USA},
  \bibinfo{pages}{14--26}.
\newblock
\showISBNx{978-1-4503-6977-0}
\urldef\tempurl%
\url{https://doi.org/10.1145/3357390.3361024}
\showDOI{\tempurl}


\bibitem[\protect\citeauthoryear{Ottoni}{Ottoni}{2018}]%
        {10.1145/3192366.3192374}
\bibfield{author}{\bibinfo{person}{Guilherme Ottoni}.}
  \bibinfo{year}{2018}\natexlab{}.
\newblock \showarticletitle{HHVM JIT: A Profile-Guided, Region-Based Compiler
  for PHP and Hack}. In \bibinfo{booktitle}{\emph{Proceedings of the 39th ACM
  SIGPLAN Conference on Programming Language Design and Implementation}}
  (Philadelphia, PA, USA) \emph{(\bibinfo{series}{PLDI '18})}.
  \bibinfo{publisher}{Association for Computing Machinery},
  \bibinfo{address}{New York, NY, USA}, \bibinfo{pages}{151–--165}.
\newblock
\showISBNx{9781450356985}
\urldef\tempurl%
\url{https://doi.org/10.1145/3192366.3192374}
\showDOI{\tempurl}


\bibitem[\protect\citeauthoryear{Paleczny, Vick, and Click}{Paleczny
  et~al\mbox{.}}{2001}]%
        {paleczny2001java}
\bibfield{author}{\bibinfo{person}{Michael Paleczny},
  \bibinfo{person}{Christopher Vick}, {and} \bibinfo{person}{Cliff Click}.}
  \bibinfo{year}{2001}\natexlab{}.
\newblock \showarticletitle{{The Java HotspotTM Server Compiler}}. In
  \bibinfo{booktitle}{\emph{Proceedings of the 2001 Symposium on JavaTM Virtual
  Machine Research and Technology Symposium - Volume 1}} (Monterey, California)
  \emph{(\bibinfo{series}{JVM'01})}. \bibinfo{publisher}{USENIX Association},
  \bibinfo{address}{USA}, \bibinfo{pages}{1}.
\newblock


\bibitem[\protect\citeauthoryear{Pall}{Pall}{2005}]%
        {luajit}
\bibfield{author}{\bibinfo{person}{Mike Pall}.}
  \bibinfo{year}{2005}\natexlab{}.
\newblock \bibinfo{title}{{A Just-in-time Compiler for Lua Programming
  Language}}.
\newblock
\newblock
\urldef\tempurl%
\url{http://luajit.org/index.html}
\showURL{%
\tempurl}


\bibitem[\protect\citeauthoryear{Rigger, Grimmer, Wimmer, W\"{u}rthinger, and
  M\"{o}ssenb\"{o}ck}{Rigger et~al\mbox{.}}{2016}]%
        {Rigger:2016:BLL:2998415.2998416}
\bibfield{author}{\bibinfo{person}{Manuel Rigger}, \bibinfo{person}{Matthias
  Grimmer}, \bibinfo{person}{Christian Wimmer}, \bibinfo{person}{Thomas
  W\"{u}rthinger}, {and} \bibinfo{person}{Hanspeter M\"{o}ssenb\"{o}ck}.}
  \bibinfo{year}{2016}\natexlab{}.
\newblock \showarticletitle{Bringing Low-level Languages to the JVM: Efficient
  Execution of LLVM IR on Truffle}. In \bibinfo{booktitle}{\emph{Proceedings of
  the 8th International Workshop on Virtual Machines and Intermediate
  Languages}} (Amsterdam, Netherlands) \emph{(\bibinfo{series}{VMIL '16})}.
  \bibinfo{publisher}{ACM}, \bibinfo{address}{New York, NY, USA},
  \bibinfo{pages}{6--15}.
\newblock
\showISBNx{978-1-4503-4645-0}
\urldef\tempurl%
\url{https://doi.org/10.1145/2998415.2998416}
\showDOI{\tempurl}


\bibitem[\protect\citeauthoryear{Rigo, Fijalkowski, and et~al.}{Rigo
  et~al\mbox{.}}{2020}]%
        {PyPyInterp}
\bibfield{author}{\bibinfo{person}{Armin Rigo}, \bibinfo{person}{Maciej
  Fijalkowski}, {and} \bibinfo{person}{Carl Friedrich Bolz-Tereick et al.}}
  \bibinfo{year}{2020}\natexlab{}.
\newblock \bibinfo{title}{PyPy's Interpreter Definition}.
\newblock
\newblock
\newblock
\shownote{\url{https://foss.heptapod.net/pypy/pypy/-/blob/branch/default/pypy/interpreter/pyopcode.py\#L1214},
  visited 2020-09-07.}


\bibitem[\protect\citeauthoryear{Sumii}{Sumii}{2005}]%
        {Sumii2005}
\bibfield{author}{\bibinfo{person}{Eijiro Sumii}.}
  \bibinfo{year}{2005}\natexlab{}.
\newblock \showarticletitle{{MinCaml: A Simple and Efficient Compiler for a
  Minimal Functional Language}}.
\newblock \bibinfo{journal}{\emph{FDPE: Workshop on Functional and Declaritive
  Programming in Education}}, \bibinfo{pages}{27--38}.
\newblock
\showISBNx{1595930671}
\urldef\tempurl%
\url{https://doi.org/10.1145/1085114.1085122}
\showDOI{\tempurl}


\newpage


\bibitem[\protect\citeauthoryear{Team}{Team}{1987}]%
        {gcc}
\bibfield{author}{\bibinfo{person}{GCC Team}.} \bibinfo{year}{1987}\natexlab{}.
\newblock \bibinfo{title}{GCC, the GNU Compiler Collection}.
\newblock
\newblock
\urldef\tempurl%
\url{https://gcc.gnu.org/}
\showURL{%
\tempurl}


\bibitem[\protect\citeauthoryear{{Topaz Project}}{{Topaz Project}}{2012}]%
        {TopazInterp}
\bibfield{author}{\bibinfo{person}{{Topaz Project}}.}
  \bibinfo{year}{2012}\natexlab{}.
\newblock \bibinfo{title}{Topaz's Interpreter Definition}.
\newblock
\newblock
\newblock
\shownote{\url{https://github.com/topazproject/topaz/blob/master/topaz/frame.py\#L141},
  visited 2020-09-07.}


\bibitem[\protect\citeauthoryear{Ungar and Smith}{Ungar and Smith}{1987}]%
        {SELF:10.1145/38807.38828}
\bibfield{author}{\bibinfo{person}{David Ungar} {and}
  \bibinfo{person}{Randall~B. Smith}.} \bibinfo{year}{1987}\natexlab{}.
\newblock \showarticletitle{Self: The Power of Simplicity}. In
  \bibinfo{booktitle}{\emph{Conference Proceedings on Object-Oriented
  Programming Systems, Languages and Applications}} (Orlando, Florida, USA)
  \emph{(\bibinfo{series}{OOPSLA '87})}. \bibinfo{publisher}{Association for
  Computing Machinery}, \bibinfo{address}{New York, NY, USA},
  \bibinfo{pages}{227–242}.
\newblock
\showISBNx{0897912470}
\urldef\tempurl%
\url{https://doi.org/10.1145/38765.38828}
\showDOI{\tempurl}


\bibitem[\protect\citeauthoryear{W\"{u}rthinger, W\"{o}undefined, Stadler,
  Duboscq, Simon, and Wimmer}{W\"{u}rthinger et~al\mbox{.}}{2012}]%
        {Wurthinger2012}
\bibfield{author}{\bibinfo{person}{Thomas W\"{u}rthinger},
  \bibinfo{person}{Andreas W\"{o}undefined}, \bibinfo{person}{Lukas Stadler},
  \bibinfo{person}{Gilles Duboscq}, \bibinfo{person}{Doug Simon}, {and}
  \bibinfo{person}{Christian Wimmer}.} \bibinfo{year}{2012}\natexlab{}.
\newblock \showarticletitle{Self-Optimizing AST Interpreters}. In
  \bibinfo{booktitle}{\emph{Proceedings of the 8th Symposium on Dynamic
  Languages}} (Tucson, Arizona, USA) \emph{(\bibinfo{series}{DLS '12})}.
  \bibinfo{publisher}{Association for Computing Machinery},
  \bibinfo{address}{New York, NY, USA}, \bibinfo{pages}{73–82}.
\newblock
\showISBNx{9781450315647}
\urldef\tempurl%
\url{https://doi.org/10.1145/2384577.2384587}
\showDOI{\tempurl}


\end{thebibliography}



\end{document}